\documentclass[useAMS,usenatbib,usegraphicx]{mn2e}
\usepackage[utf8]{inputenc}
\usepackage[usenames]{color}
\usepackage[dvipsnames]{xcolor}
\usepackage{ulem}
\usepackage{graphicx}   
\usepackage[T1]{fontenc}
\usepackage{ae,aecompl}
\usepackage{hyperref}

\usepackage[T1]{fontenc}
\usepackage{ae,aecompl}

\usepackage{graphicx}	
\usepackage{amsmath}	
\usepackage{amssymb}	

\def\apss{Ap\& SS}
\def\apj{ApJ}
\def\apjl{ApJL}
\def\aap{A\& A}

\def\araa{ARA\&A}
\def\mnras{MNRAS}

\def\aj{{AJ}}
\def\apjs{ApJS}

\def\mnras{{MNRAS}}

\def\rmxaa{Rev. Mex. Astron y Astrofis.}
\def\ssr{Space Sci. Rev.}
\def\aapr{Astron. \&\ Astrophys. Rev`}

\def\alamenos#1{$^{-#1}$}

\newcommand{\alphavir}{\alpha_{\rm vir}} 
\newcommand{\alphavirBM}{\alpha_{\rm vir}^{\rm BM92}} 
\newcommand{\alphavirfull}{\alpha_{\rm vir}^{\rm full}}

\newcommand{\alphavirclass}{\alpha_{\rm vir}^{\rm class}} 
\newcommand{\alphavirtyp}{\alpha_{\rm vir}^{\rm class}}

\newcommand{\cc}{{\rm cm^{-3}}}

\newcommand{\Egrav}{E_{\rm g}}    %
\newcommand{\Egravcl}{E_{\rm g, cl}}    %
\newcommand{\Egravsph}{E_{\rm g, \circ}}    %
\newcommand{\Ekin}{E_{\rm k}}    %

\newcommand{\kms}{{\rm km\ s^{-1}}}    %

\newcommand{\Phicl}{\Phi_{\rm cl}}    %

\newcommand{\Reqsph}{R_{\rm eq, \circ}}

\newcommand{\tauff}{\tau_{\rm ff}}

\title[Why most molecular clouds are gravitationally dominated]{Why most molecular clouds are gravitationally dominated}

\author[L. Ram\'irez-Galeano, J. Ballesteros-Paredes, R. Smith, et al]{
Laura Ramírez-Galeano$^{1}$\thanks{E-mail: lauramirezg26@gmail.com}, Javier Ballesteros-Paredes$^{2}$,  Rowan J. Smith$^3$, \newauthor
Vianey Camacho$^{4}$ \& Manuel Zamora-Avilés$^4$
\\
\\
$^1$Instituto de F\'isica, Universidad de Antioquia, Calle. 67 No. 53-108, Medell\'in, Colombia
\\
$^{2}$Instituto de Radioastronom\'ia y Astrof\'isica, UNAM, campus Morelia. PO Box 3-72. 58090. Morelia, Michoac\'an, M\'exico
\\
$^3$Jodrell Bank Centre for Astrophysics, Department of Physics and Astronomy, University of Manchester, Oxford Road, Manchester M13 9PL, UK
\\
$^{4}$Instituto Nacional de Astrof{\'i}sica, {\'O}ptica y Electr{\'o}nica, Luis E. Erro 1, 72840 Tonantzintla, Puebla, M{\'e}xico
}

\date{Accepted XXX. Received YYY; in original form ZZZ}

\pubyear{2015}

\begin{document}
\label{firstpage}
\pagerange{\pageref{firstpage}--\pageref{lastpage}}
\maketitle

\begin{abstract}

Observational and theoretical evidence suggests that a substantial population of molecular clouds (MCs) appear to be unbound,  {\color{black}dominated by turbulent motions.} However, these estimations are made typically via the so-called viral parameter $\alphavirclass$, which is an observational proxy to the virial ratio between the kinetic and the gravitational energy. This parameter intrinsically assumes that MCs are isolated, spherical, and with constant density. {\color{black}  However, MCs are embedded in their parent galaxy and thus are subject to compressive and disruptive tidal forces from their galaxy, exhibit irregular shapes, and show substantial substructure.}  We, therefore, compare the typical estimations of $\alphavirclass$ to a more precise definition of the virial parameter, $\alphavirfull$, which {\color{black} accounts not only for the self-gravity (as $\alphavirclass$), but also for the tidal stresses, and thus, it can take negative (self-gravity) and positive (tides) values.} While we recover the classical result that most of the clouds appear to be unbound, having $\alphavirclass > 2$, we show that, with the more detailed definition considering the full gravitational energy, {\color{black} (i) 50\%\ of the total population is gravitationally bound, however, (ii) another 20\%\ is gravitationally dominated, but with tides tearing them apart; (iii) the source of those tides does not come from the galactic structure (bulge, halo, spiral arms), but from the molecular cloud complexes in which clouds reside, and probably (iv) from massive young stellar complexes, if they were present. (v) Finally, our results also suggest that, interstellar turbulence can have, at least partially, a gravitational origin}.  
\end{abstract}

\begin{keywords}
turbulence -- stars: formation -- ISM: clouds -- ISM: kinematics and dynamics -- galaxies: star formation.
\end{keywords}



\section{Introduction}\label{sec:intro}

Molecular clouds (MCs) are the coldest and densest sites of the interstellar medium and, consequently, the stars' birthplace. Their masses vary over many orders of magnitude, but in general, the great majority of them are substantially larger than the so-called Jeans mass, i.e., the mass that thermal energy can support against collapse, 
\begin{equation}
    M_J \sim 2 M_\odot \bigg(\frac{c_s}{0.2\ \kms}\bigg)^3 \bigg( \frac{n}{10^3\ \cc}  \bigg)^{-1/2},
    \label{eq:JeansMass}
\end{equation}
where $c_{s}$ is the sound speed, and $n$ is the number density. At face value, thus, MCs should be collapsing within a few times  free-fall timescales 
{\color{black}
\begin{equation}
    \tauff = 3.4~\mathrm{Myr}~ \bigg(\frac{n}{100~\mathrm{cm}^{-3}}\bigg)^{-1/2},
    \label{eq:tauff}
\end{equation}
}
\citep{Galvan-Madrid+07}, where $G$ is the constant of gravity and $\rho$ the density, (e.g., with $\tauff\sim 3.5$~Myr for an MC with a mean density of 100~cm\alamenos 3). If clouds were collapsing monolithically, the star formation efficiency, computed as the mass in newborn stars in a given MC over the mass of the MC, should be as large as 50\%\ or more. However, typical values of the efficiency are around a few 1\%\ \citep[e.g.,][]{Johnstone+04, Enoch+06, Young+06}, suggesting either (a)~that clouds are supported by other mechanisms in addition to thermal pressure \citep{Shu+87, MacLowKlessen04, Ballesteros-Paredes+07, McKeeOstriker07}, (b)~that the stars are formed rapidly and they destroy their parent cloud before the star formation efficiency increases substantially \citep[e.g., ][]{Ballesteros-Paredes+99b, Ballesteros-ParedesHartmann07, VazquezSemadeni+07}, or (c)~that during the process of star formation, clouds continue increasing their mass as the instantaneous rate of star formation increases, keeping the efficiency with small values before the clouds are destroyed by UV radiation feedback from their massive stars \citep[e.g.,][]{VazquezSemadeni+10, ZamoraAviles+12, Hartmann+12}. 

Among these possibilities, the currently more supported scenario of molecular cloud dynamics and star formation is the first one: that clouds are globally supported by turbulence and magnetic fields against gravity, and that collapse occurs only in small regions where turbulent motions gather enough mass together to become gravitationally unstable \citep[see, e.g., ][and references therein]{MacLowKlessen04, Ballesteros-Paredes+07}. 

The idea of clouds being supported by turbulence has existed for more than 70 years. \citet[][]{Chandra51} suggested that the Jeans mass can be modified by replacing the sound speed $c_s$ in eq. (\ref{eq:JeansMass}) by an effective sound speed, given by 
\begin{equation}
    c^2_{s, \rm eff} = c_s^2 + \frac{1}{3} \sigma_v^2 ,
\end{equation}
where $\sigma_v$ is the 3D non-thermal velocity dispersion of the gas. {It should be noticed, however, that this approach suggests that turbulent motions are statistically isotropic, and play at the smaller scales. If this is not the case, then rather than preventing collapse, they may very well promote it \citep{Ballesteros-Paredes+99a, Ballesteros-Paredes06}}.

For several decades, it has been thought that one way to distinguish whether an MC is dominated by gravity or by non-thermal motions is by evaluating the relative importance between the kinetic and gravitational energies \citep[e.g., ][]{Larson81, Carr87, Loren89, BertoldiMcKee92, Kauffmann+13}. Observationally, the available variables are the mass, velocity dispersion and size. This lead to \citet{BertoldiMcKee92} to define the so-called virial parameter (which is nothing but a proxy to the virial ratio between the kinetic and gravitational energies) as
\begin{equation}
    \alphavirBM \equiv \frac{5 \sigma^2_{v, \rm 1D} R}{GM} , 
    \label{eq:alphavirBM}
\end{equation}
where $G$ is the constant of gravity, and $R$, $M$ and $\sigma_{v, \rm 1D}$ the size, mass, and 1D velocity dispersion of the cloud.

Although former estimations of these quantities for MCs indicated $\alphavirBM\sim 1$, i.e., that gravity is in some sort of equipartition with the kinetic energy \citep[e.g., ][]{Larson81, Solomon+87, MyersGoodman88}, further estimations showed an anti-correlation between $\alphavir$ and the mass of the cloud, such that $\alphavirBM\sim 1$ occurs only for the largest clouds of each sample, while most clouds exhibit supervirial values \citep{Loren89, BertoldiMcKee92, Kauffmann+13, Miville-Deschenes+17}, frequently by more than one order of magnitude.

{\color{black}
The fact that there is a population of overvirial clouds (the less massive ones in each survey)  has} several interesting implications. On one hand,  \citet{Field+11} argued that those supervirial clouds and cores could be interpreted as pressure confined. The problem with this approach is that the pressures of the interstellar medium required to confine clouds vary between $P/k\sim 10^4$~K~cm\alamenos3 and $P/k\sim 10^7$~K~cm\alamenos3 \citep[see Fig.~3 in][]{Field+11}. Such pressures are always substantially larger than the characteristic pressure of the ISM $P/k\sim5\times 10^3$~K cm\alamenos 3 \citep{Elmegreen89}. Second, virial equilibrium (i.e., $\ddot I=0$, with $I$ the moment of inertia of the cloud) and pressure confinement, are two conditions hard to reach in dynamical MCs  \citep{Ballesteros-Paredes+99a, Ballesteros-Paredes+07}.

Another possibility is whether the anti-correlation could be an artifact of observational methods. For instance, \citet[][]{Traficante+18} argued that the anti-correlation could be due to the fact that the emitting volumes from which we infer the mass (dust continuum) and the velocity dispersion (line emission) in massive cores are different.  Although interesting, this possibility has two caveats: (i)~it makes use of spherical symmetry, while the anti-correlation occurs in clouds, clumps, and cores that, in general, are far from spherical, (ii)~the anti-correlation appears also in observations where the mass is not necessarily inferred from dust continuum, but also line emission \citep[e.g.,][]{Miville-Deschenes+17}, and thus, the hypothesis that the emission volumes are different is not valid.

A third interpretation for the anti-correlation between $\alphavirBM$ and the mass is that, rather than an artifact of the observational methods, it is an artifact of our definition of clouds. Indeed, when defining clouds through a single column density threshold, a reasonably well-defined, though artificial, mass-size power-law $M\propto R^p$ naturally arises \citep[][]{Ballesteros-Paredes+12}. Such a relation, in combination with a substantially weaker correlation between the velocity dispersion and size, $\sigma_v\propto R^q$, produces the $\alphavirBM$ {\it vs.} $M$ anti-correlation \citep[][]{Kauffmann+13, Ballesteros-Paredes+20}.

Overvirial clouds have been found also in numerical simulations of molecular clouds \citep[e.g., ][]{Dobbs+11, Duarte-CabralDobbs16, Padoan+16, Tress+21, Ganguly+22}, and the excess of kinetic energy has been interpreted as a sign that most MCs are unbound, and probably under pressure confinement. 

It should be noticed that there is a contradiction, however, between the estimations of $\alphavirBM$ and the standard picture of cloud dynamics, namely, large clouds being unbound and only cores becoming self-gravitating and collapsing. At face value, $\alphavirBM<1$ occurs only for the largest clouds \citep[see, e.g., Fig 8, 3 and 1 in][respectively]{Loren89, BertoldiMcKee92, Kauffmann+13}. These are the clouds that should collapse since these are the ones that are mostly bound. In contrast, smaller clouds or clumps have large values of $\alphavirBM$, and then, should be unbound. 

In order to solve this contradicction, we notice that $\alphavirBM$intrinsically assumes that the graivitational energy can be approached by the gravitational energy of an isolated, homogeneous sphere:
\begin{equation}
    \Egravsph = - \frac{3}{5} \frac{G M^2}{R}, 
    \label{eq:Egravsph}
\end{equation}
where $G$ is the constant of gravity, $M$ is the mass of the cloud, and $R$ is an effective radius, typically computed as $R=\sqrt{A/\pi}$, with $A$ the projected area of the cloud in the sky. However, there are several caveats to this approach: 
\begin{enumerate}

    \item{} Eq.~(\ref{eq:Egravsph}) is just a lower limit (in absolute value) of the actual gravitational content since it does not consider the inner structure of the clouds. In fact, estimations of the gravitational content of MC cores in simulations and in a few observed cores, \citep{Ballesteros-Paredes+18}, showed that the actual gravitational energy of cores with inner structure could be substantially larger, in absolute value, than the gravitational energy as computed by eq.~(\ref{eq:Egravsph}).    
    \item{} In addition to self-gravity, clouds are not isolated: they respond to the total gravitational potential of their host galaxy, and tidal effects can be part of their total gravitational budget \citep{BonnellRice08, Ballesteros-Paredes+09a, Ballesteros-Paredes+09b, Suarez+12, RomeoFalstad13, Jog13, Renaud+14, NtormousiHennebelle15, Baba+17, Meidt+18, Singh+19, Dale+19}. In fact, half the population of clouds in numerical simulations may have converging motions even if they have large virial parameters \citep{Vazquez-Semadeni+08, Camacho+16, Baba+17}.

    \end{enumerate}

{\color{black}
{Assuming that turbulent motions are at the smaller scales only, and they all provide support, in order to estimate the actual dynamical state of MCs, it still } becomes necessary to modify the virial parameter, such that the actual total gravitational potential is taken into account. This involves the stellar disk, stellar spiral arms, bulge, bar, dark matter halo, sink particles, and the gas itself.  At first approximation, the gravitational potential {\color{black} of the dark matter halo, which gives rise to the circular} rotation curve can be considered negligible. Indeed, as shown by \citet[][]{Mihalas_Routhly68} and \citet{Jog13}, in the Solar Neighborhood, assuming spherical symmetry, structures with densities larger than 
\begin{equation}
  \rho_{\mathrm{crit}} > \frac{3}{2 \pi G} \ A (A-B) \sim \mathrm{few\ 10^{-24} ~cm^{-3}}
  \label{eq:rocrit_tides}
\end{equation}
(where $A$ and $B$ are the Oort constants) will not be torn apart by the differential rotation because its self-gravity dominates. This suggests that molecular clouds are not strongly subject to galactic tidal forces. Similarly, \citet{Suarez+12} showed that for a typical bulge and/or extended halo, the gravitational pull by these potentials over molecular clouds modeled as Plummer spheroids is indeed negligible.  Note, however, that this may not be the case if the geometry of the cloud is different and if it includes spiral arms. In fact, it is clear that somehow the gas responds to the stellar spiral arms by enhancing its density and increasing its star formation. Thus, the detailed position, orientation and size of the cloud with respect to the spiral pattern may play a role in the energy budget of the cloud \citep{Ballesteros-Paredes+09a, Ballesteros-Paredes+09b}.
}

In this work, using numerical simulations of a Milky Way-type galaxy from the suit called `The Cloud Factory' \citep{Smith+20} we study the whole gravitational content of molecular clouds and how their inner structure (fractal shape, non-constant density), as well as the tides from nearby clouds, affect their energy budget. {\color{black} In \S\ref{sec:GravitationalContent} we define the virial parameter in two different ways: as it was defined by \citet{BertoldiMcKee92}, and by considering the actual gravitational forces through the cloud, i.e.,} without making any assumptions about the geometry of the cloud or mass distribution.  {\color{black}  In \S\ref{sec:numerical} we briefly summarize the simulations from \citet{Smith+20} and the usage of the cloud extraction algorithm \citep[][]{Camacho+16}. In addition, we also explain how we compute the gravitational forces involved, since the present work is a post-processing analysis of already performed simulations, and no information on the gravitational forces, was previously stored. } In  \S\ref{sec:results} and \S\ref{sec:discussion} we present our results and the corresponding discussion on the actual energy budget of molecular clouds. Finally, we summarize our conclusions in \S\ref{sec:conclusions}.

\section{Gravitational content of MCs}\label{sec:GravitationalContent}

The actual gravitational content $W$ of any parcel of fluid  arises in the virial theorem by taking the dot product of the gravitational force $\rho~\partial\Phi/\partial x_i$ by the position vector $x_i$  and integrating over volume \citep[e.g., ][]{Parker79, Shu92, MZ92},
\begin{equation}
    W = - \int_V x_i \rho \frac{\partial\Phi}{\partial x_i} dV, 
    \label{eq:W}
\end{equation}
{\color{black}
where we have used the Einstein summation convention over repeated indexes. } This term accounts for the whole gravity of the system over the volume $V$, since $\Phi$ is the total gravitational potential, {\color{black} which is due to all the mass contributing to the total gravitational force of the system. In addition, since we are interested in computing the energies in the frame of reference of the cloud, which revolves around the galaxy, it becomes necessary to include also the centrifugal and Coriolis forces, given by  $\rho\pmb{\omega} ( \pmb{\omega}\times\pmb{r})$ and $2\rho\pmb{\omega}\times\pmb{\upsilon}$, respectively \citep[see, e.g., ][]{Ballesteros-Paredes+09a, Baba+17}, where $\pmb{\omega}$ is the angular frequency of the rotation of the frame of reference of the cloud, and $\pmb{r}$ and $\pmb{\upsilon}$ are the position and velocity vectors of a given element of volume in the cloud. Thus, the total ``effective'' gravitational term in the rotating frame will include the components from the halo (h), bulge (b), stellar disk (d), stellar spiral arms (a), centrifugal (cen), Coriolis (Cor), gas (g) and sink particles (s), the latter ones representing young stellar clusters in our simulations, i.e., 

\begin{equation}
 W_{\mathrm{tot}} = W_h + W_b + W_d + W_a - W_\mathrm{cen} - W_\mathrm{Cor} + W_g + W_s
 \label{eq:W_sum}
\end{equation}
}
Assuming now that the cloud is isolated, ($\Phi = \Phicl$), $W_\mathrm{tot}$ can be rewritten as the gravitational energy of the cloud \citep[see, e.g., ][]{Shu92}, $\Egravcl$
\begin{equation}
    \Egravcl = - \frac{1}{2} \int_V \rho \Phicl\ dV = 
  -\frac{1}{2}G \int_V  \int_V \frac{\rho(\mathbf{r}) \rho(\mathbf{r'})   }{|\mathbf{r}-\mathbf{r'} |} \mathrm{d}^3r\ \mathrm{d}^3r'
    \label{eq:Egravcl}
\end{equation}
where we have now explicitly added the subscript {$_\mathrm{cl}$} to denote that the gravitational potential is only the one produced by the mass within the volume $V$. Although clouds exhibit highly filamentary morphologies \citep[e.g., ][]{Andre+14}, $\Egravcl$ is frequently computed as if clouds were spheres with constant density, and thus, $\Egravcl$ is approached as $\Egravsph$ in eq. (\ref{eq:Egravsph}). 

In order to understand the differences {\color{black} between an isolated, homogeneous, spherical cloud and a cloud with inner structure embedded in an external potential, we define the $\alphavir$} in two different ways. The first one, which we call the classical virial alpha, $\alphavirclass$, will be computed in terms of the kinetic to gravitational energy ratio, which in principle, is equivalent to the observational definition of $\alphavir$ by \citet{BertoldiMcKee92}\footnote{We distinguish $\alphavirclass$ from $\alphavirBM$ because the latter is defined in terms of observed quantities (particularly, the 1D velocity dispersion), while with the former, we use the whole 3D data from the simulations.}, 
\begin{equation}
    \alphavirclass \equiv \frac{2 \Ekin}{|\Egravsph|} \sim \alphavirBM .
    \label{eq:alphavirclass}
\end{equation}
The second one, which we call the full virial parameter, $\alphavirfull$, 
{\color{black}  is obtained by replacing $\Egravsph$ in eq.~(\ref{eq:alphavirclass}) by the term $W$ provided by eq.~(\ref{eq:W}):
\begin{equation}
    \alphavirfull \equiv \frac{2 \Ekin}{W}.
    \label{eq:alphavirreal}
\end{equation}
In other words, the gravitational potential that it is used now is the total gravitational potential due to all possible agents that could be playing a gravitational role over the volume $V$.}

A first important difference is that, in contrast to $\alphavirclass$, $\alphavirfull$ inherits the sign of $W$, and thus, it can be either positive or negative. If $W>0$, the gravitational term contributes to the disruption of the cloud, i.e., tidal forces overcome the gravitational energy of the cloud.  If negative, it contributes to the collapse of the cloud. 

\begin{figure}
     \center{\includegraphics[width=0.8\columnwidth]{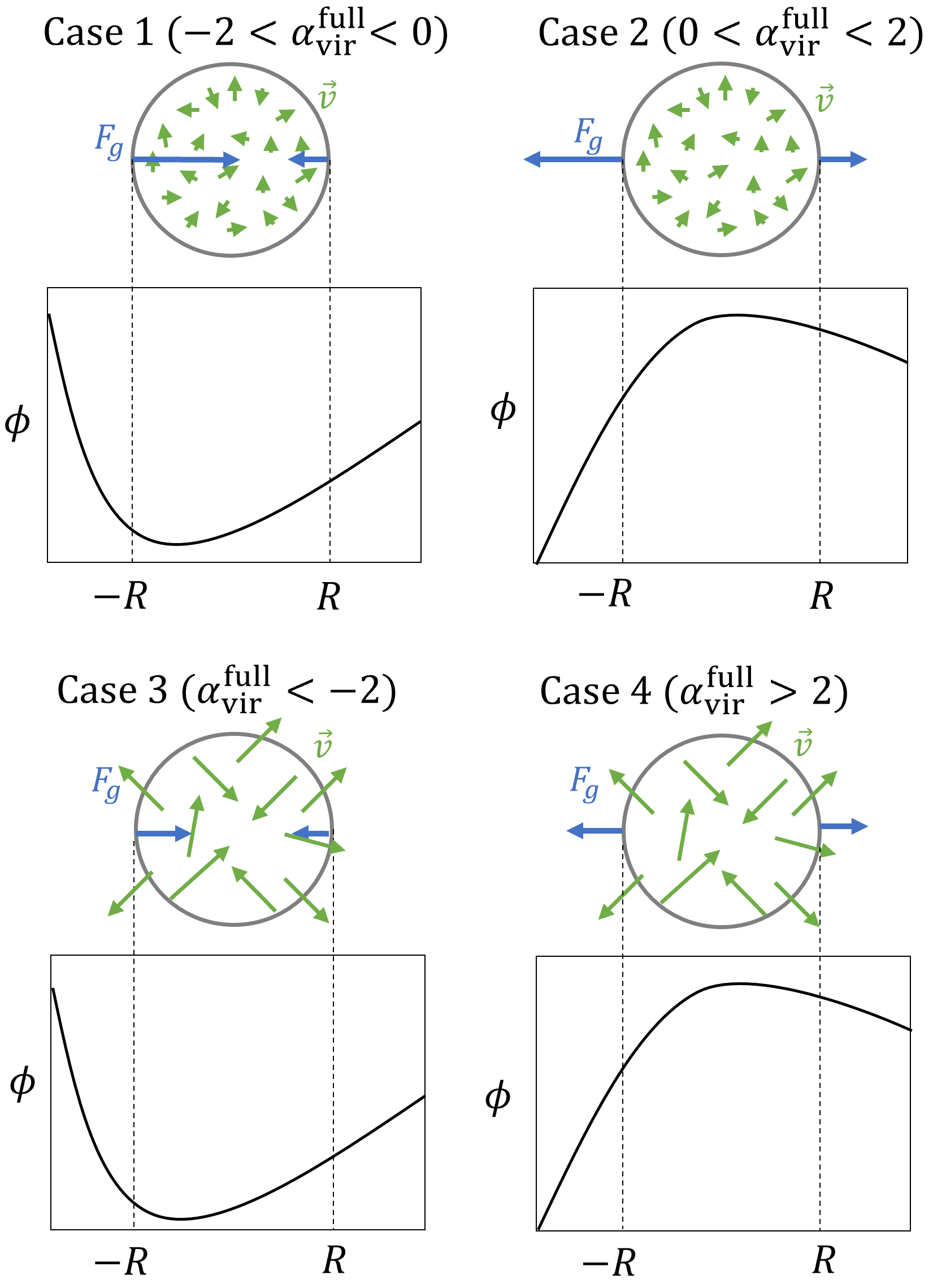}}
    \caption{{Diagram for a toy cloud. The cloud is represented by a circle, the gravitational potential is outlined in the boxes below the cloud, blue arrows represent the gravitational force due to the potential gradient, and green arrows represent the velocity dispersion within the cloud. We show in four cases the different ranges of values that the virial parameter can obtain as a consequence of the behavior of the velocity dispersion and gravitational potential. In cases, 1 and 3 the gravitational potential has a positive concavity so the total gravity force has a compressive role (i.e $W\leq0$), while in cases 2 and 4 the tidal forces contribute to tearing apart the cloud (i.e $W>0$).}}
    \label{fig:example}
\end{figure}

{\color{black}
For the analysis of the present work, we will consider a cloud with $E_k + E_g < 0$ as gravity-bound in the classical analysis, and with $E_k + W_\mathrm{tot} < 0$ in the new formalism. While in the first case, the condition will imply that $\alphavirclass<2$, the second condition requires to reconsider the sign, and thus, $-2 < \alphavirfull < 0$. For symmetry, we assume that if tidal stresses dominate over turbulence, $0 < \alphavirfull<2$.} In order to understand the physical meaning of $\alphavirfull$, in Fig.~\ref{fig:example} we draw schematically four possible cases of a toy cloud (round cloud), the velocity dispersion within the cloud (green arrows), and the total gravitational force (blue arrows) due to the gradient of the (total) gravitational potential $\Phi$, shown in the plot below the drawing. In the upper panels, the gravitational energy overwhelms the kinetic energy (though note this does not lead to collapse in the case of the top-right system), and thus, $-2 < \alphavirfull < 2$. This situation is represented schematically by large gravitational force arrows and small velocity arrows.  In the lower panels, the situation is reversed: the kinetic energy is larger than the gravitational energy ($2 < |\alphavirfull|$), which is cartooned by larger velocity vectors and small gravitational force vectors. By columns, the left panels of Fig.~\ref{fig:example} indicate a situation in which the total gravitational potential has concavity pointing upwards, and thus, the total gravity of the system is compressive, as cartooned by the blue converging arrows. In the right-hand panels, this situation is reversed: the total gravitational potential has concavity downwards, which means that the gravity is disruptive. Case by case, when $-2 < \alphavirfull<0$ (upper-left panel), gravity is compressive and wins over turbulence. The cloud should collapse if no other forces are present. If instead, $0< \alphavirfull < 2$ (upper-right panel), again gravity wins, but it is disruptive, since $W>0$. The cloud is torn apart by tidal forces. If $\alphavirfull< -2$ (lower-left panel), kinetic energy wins over a net compressive gravitational energy. Such a system will expand. Finally, if $\alphavirfull> 2$ (lower-right panel), the gravitational and kinetic energy are both disruptive. The cloud is torn apart {mainly} by turbulent motions.

\section{Numerical data and post-processing}\label{sec:numerical}

\subsection{Numerical simulations}

In order to understand the influence of the total gravitational potential on MCs, we use numerical data from the ``Cloud Factory'' Suite \citep[][]{Smith+20}, which uses a modified version of the AREPO code \citep{Springel10}, that incorporates a Galactic potential with a spheroidal bulge, a thin $+$ thick stellar disk, a \citet[][]{Navarro+96} dark matter halo, and a spiral perturbation based on the prescription by \citet[][]{CoxGomez02}. In addition, the code also includes time-dependent gas chemistry and self-shielding, cooling, self-gravity, star formation via the formation of sink particles, and stellar feedback through supernova explosions. Details on the numerical code can be found in \citet[][]{Smith+14,  Smith+20}, and references therein.

In the present work we focused on ``Region A'', a $ 200 \times 200\times 200$ parsec region of the potential-dominated simulation performed by \citet[][see their Fig.~1, upper panels]{Smith+20}. In this simulation, the stellar feedback was random before self-gravity was turned on, and thus, the supernovae were inefficient at pushing the dense gas, allowing the creation of well-defined gas spiral arms. We analyzed this simulation at 2 Myr after self-gravity was turned on. 

The sink particles are non-gaseous particles that allow us to mimic star formation sites. They are introduced (a) in places where the gas density exceeds a critical density and (b) satisfy a series of energy checks to ensure that the gas within their radius is unambiguously bound and collapsing  \citep[for details, see][\S~2.4]{Smith+20}.  We consider them in the evaluation of the gravitational potential because they arise from the gas distribution of the cloud, and consequently, their gravitational contribution to embedded and adjacent molecular clouds could be relevant.

 \subsection{Finding clouds in the simulation}
 \label{sec:clumpfind}
 
\begin{figure*}
    \includegraphics[width=2\columnwidth]{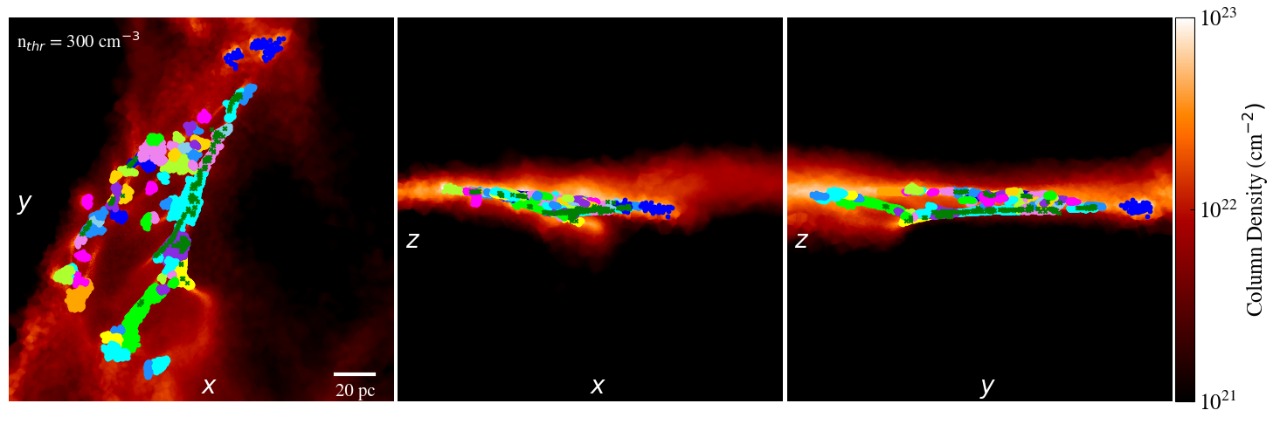}
    \caption{\color{black} Red scale: Column densities in the three projections of the analysed box. Overlaped in colors are the clouds found by our algorithm. Green dots: stellar (sink) particles.}
    \label{fig:maps}
\end{figure*}

{\color{black}
Arepo  \citep{Springel10} is a code that uses an Arbitrary Lagrangian-Eulerian (ALE) fluid dynamic method, and it is based on a moving unstructured mesh. This mesh is defined by a Voronoi tesselation of a set of ``generating points'' (GP). We used the open source {\tt voro++} routine \citep{Rycroft+09} in order to generate the Voronoi cells and compute the energies.  In addition, in order to define clumps, we make use of the GPs as if these were SPH particles,  and thus, apply the method by \citet{Camacho+16}:

\begin{enumerate}
  \item{} We first select all the GPs for which their cells have a density above a certain arbitrary threshold. 
  
  \item We then compute their characteristic length as the cubic root of the ratio between the mass and the density of the cell to which the GP belongs, i.e., 
\begin{equation}
  l_{\rm char} = \bigg(\frac{m_{\rm cell}}{\rho_{\rm cell}}\bigg)^{1/3}
  \label{eq:char_length}
\end{equation}

  \item We locate the GP with the highest density.  This particle and all those located within its characteristic length (\ref{eq:char_length}) are labeled as members of the clump. Then, the following steps are iterated: (a) we locate the member of the clump with the highest density to which this subprocedure has not been applied, and (b) we label as members all the GPs within a characteristic length not yet belonging to the clump. (c) The iteration ends when all the clump members are examined.
  
  \item If there are particles remaining with a density $n$ larger than the threshold density that are not yet members of any clump, we locate the one with the highest density and use it to define a new clump and the whole procedure is repeated. 
  
With the idea of studying only clumps that are reasonably well resolved, once the procedure has finished we rejected those clumps that have less than 40 GPs. 

\end{enumerate}

In Fig.~\ref{fig:maps} we show the three projections of our column density field (red scale). The clumps found by our algorithm are represented with the different discretized colors. We also show, with green dots, the sink particles that represent the newborn stellar clusters.}
    
{\color{black}

\subsection{Post-processing calculations}
\label{sec:ComputingForce}

With the procedure depicted in the previous section, we found 878 clumps, for which we compute the masses ($M$), sizes ($\Reqsph$), 3D velocity dispersions ($\sigma_v$), gravitational virial terms ($W$), gravitational and kinetic energies ($E_{\rm g,\circ}$ and $\Ekin$, respectively), and virial parameters ($\alphavirclass$ and $\alphavirfull$). For this purpose, we 

\begin{enumerate}

\item{} Defined also the equivalent radius of each cloud as the radius of a sphere with the same volume, i.e., %
\begin{equation}
    \Reqsph = \bigg(\frac{3}{4\pi} V_{\rm cl} \bigg)^{1/3} .
    \label{eq:Reqsph}
\end{equation}
where $V_{\rm cl}$ is the volume of the clump, defined as the sum of all cells belonging to the clump. 

\item{} In order to compute the gravitational term $W$ given by (\ref{eq:W}), we were required to compute the mean gravitational acceleration in each cell, in the non-uniform Voronoi grid. Since the simulation only stored the gravitational potential, but not the force, in order to compute the mean value of the gradient of the gravitational potential at a given position we followed  \citet[][see their eq. (21)]{Springel10}
\begin{equation}
  \langle \nabla\phi \rangle_i = \frac{1}{V_i} \sum_{j\ne i} A_{ij}
  \bigg( \big[\phi_j-\phi_i]\frac{\mathbf{c}_{ij}}{r_{ij}} - \frac{\phi_i + \phi_j}{2} \frac{\mathbf{r}_{ij}}{r_{ij}} \bigg)
  \label{eq:gradiente}
\end{equation}
where $A_{ij}$ is the area between the $i^{th}$ and $j^{th}$ neighbouring cells, $\mathbf{r}_{ij}$ is the vector pointing from the  $i^{th}$ to the $j^{th}$ GPs, and $\mathbf{c}_{ij}$ is given by 
\begin{equation}
  \mathbf{c}_{ij} \equiv \frac{1}{A_{ij}} \int_{A_{ij}} X\bigg( \mathbf{r} - \frac{r_i+r_j}{2} \bigg) dA
  \label{eq:c_ij}
\end{equation}
\citep[see][ eq.~18]{Springel10}.

\end{enumerate}

}

\section{Results}\label{sec:results}

{\color{black}

\subsection{Gravitational tides from  different components of the galaxy. }
\label{sec:GalacticTides}

Since typical studies compute the gravitational energy $\Egravsph$, but not the gravitational term $W$, we first look at the latter in order to understand the relative contribution of each component of the galaxy in the global gravitational budget of the clouds. In Fig.~\ref{fig:AllWs} we plot the total gravitational term, $W_{\mathrm{tot}}$ ($x$ axis), as defined by eq. (\ref{eq:W_sum})  {\it vs.} the individual gravitational terms, produced by, from left to right: first row, gas and sinks together (gs), the gas alone (g) and the sinks alone (s); second row, the stellar spiral arms (a), the bulge (b); and the stellar disk (d); and third row,  the dark matter halo (h), and the centrifugal (cen) and Coriolis (Cor) forces. 

\begin{figure*}
     \center{\includegraphics[width=2\columnwidth]{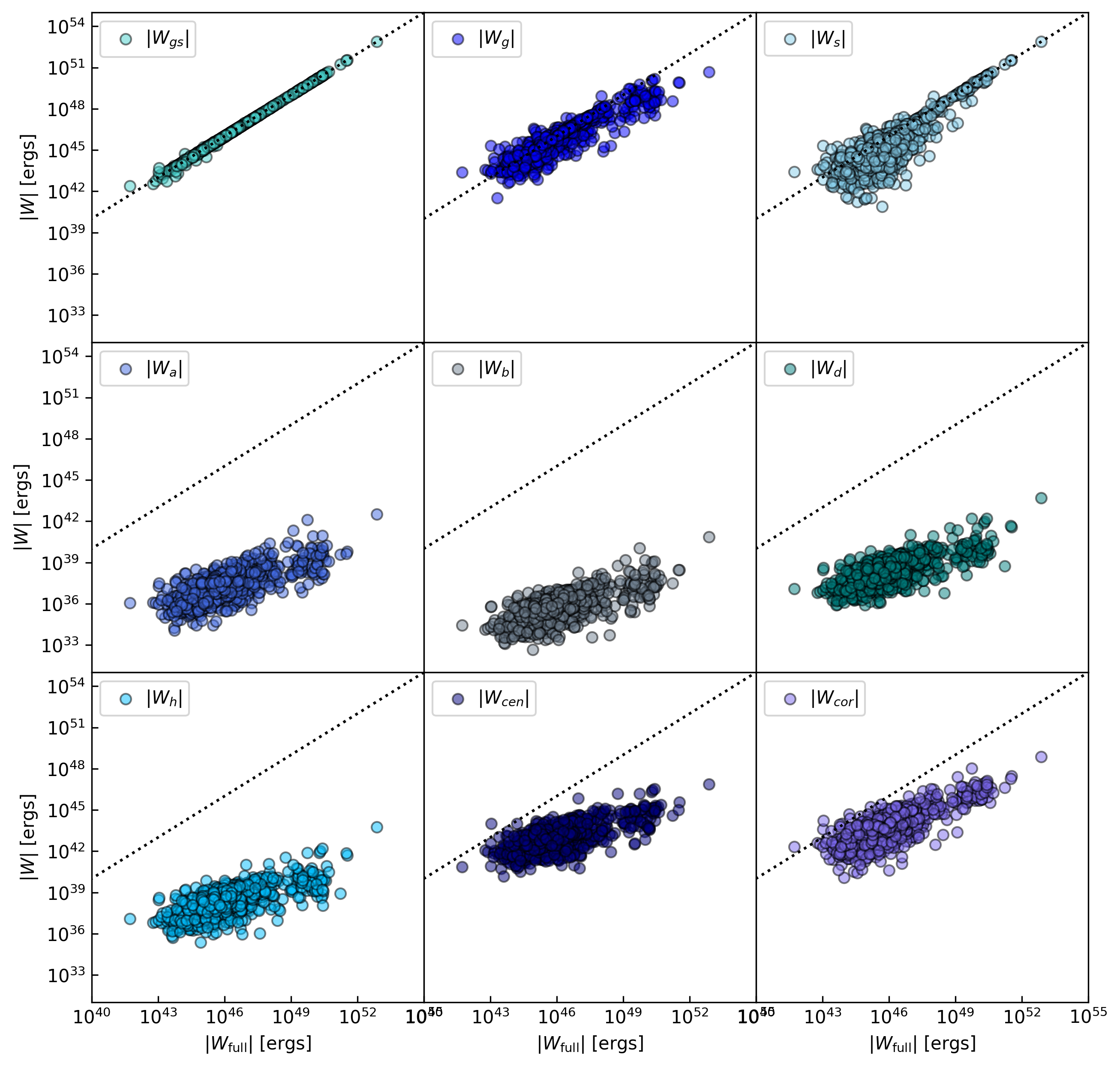} \\
  \caption{\color{black}Individual gravitational terms from the gas and sinks (gs), gas (g) and sinks (s), first row; stellar spiral arms (a), galactic bulge (b), and stellar disk (d), second row; and finally, dark matter halo (h), centrifugal (cen) and Coriolis (cor) terms, in the bottom row. The dotted line in each panel is the identity. Notice that the main contributors to the total gravitational content of MCs are the total potential from the gas itself, and the sinks. The centrifugal and Coriolis terms are 2nd order corrections, except for a few small clouds, while the galactic components are 4th order corrections.} 
  \label{fig:AllWs}
}
\end{figure*}

This plot shows that the more relevant contribution to the gravitational budget of the clouds are the terms involving the gravitational potential of the gas and of the sinks (first row). In contrast, the second and third rows of Fig.~\ref{fig:AllWs} show that the contribution from the arms, disk, bulge and halo to the total gravitational budget of the clouds is negligible. This result not only confirms the estimate of \citet{Mihalas_Routhly68, Suarez+12}, and \citet{Jog13} in the sense that the MCs are not only not strongly influenced by the spherical component of the gravitational potential, which will dominate the rotation curve of the galaxy, but it also extends the result to the stellar spiral arms and the stellar disk. 

On the other hand, the last two panels show that the corrections to be made to the gravitational term $W$ due to the centrifugal and Coriolis forces are typically more important than those from the galactic structure, although still a second-order correction in most cases, except for some clouds in the case of the Coriolis term, and a few others in the case of the centrifugal term. Between them, moreover, the Coriolis term seems to be more relevant than the centrifugal term.

In summary, the gravitational content of molecular clouds is not dominated by the soft galactic density structures (stellar spiral arms, bulge, stellar disk, and dark matter halo), but by those structures that locally exhibit the sharpest structures: gas, and sink particles. 

{
Of particular interest is the contribution from the latter. While our sinks are unresolved dense gas structures, one can imagine that at least some of them might become massive young stellar clusters, if we could follow the physics. If this were true, it may be suggesting that massive young stellar clusters could play an important role in the gravitational budget of their parent clouds. We will discuss the implications of this result in \ref{sec:clusters}.
}

\subsection{Global statistics}
\label{sec:GlobalStatistics}

\subsubsection{Gravitational term vs the gravity of a sphere with constant density}
\label{sec:WvsEgsph}

In order to estimate whether previous observational and theoretical estimates of the gravitational state of molecular clouds are adequate, we need to compare the gravitational term $W$ to the frequently used gravitational energy of the sphere with constant density, $\Egravsph$, eq.~(\ref{eq:Egravsph}). For this purpose, in Fig.~\ref{fig:histoWoverEgsph} we plot the histograms of the ratio $W/\Egravsph$. The left panel corresponds to those clouds for which $W>0$ (gravitationally stirred clouds). The middle panel includes only clouds for which $W\le0$ (gravitationally compressed clouds), while the right panel includes all the clouds found in our box. The dotted line indicates where the ratio is equal to one. The percentages in the upper part of each panel indicate the percentage of clouds in that panel for which $|W/\Egravsph|$ is smaller (left) or larger (right) than one. 

\begin{figure*}
     \center{\includegraphics[width=2\columnwidth]{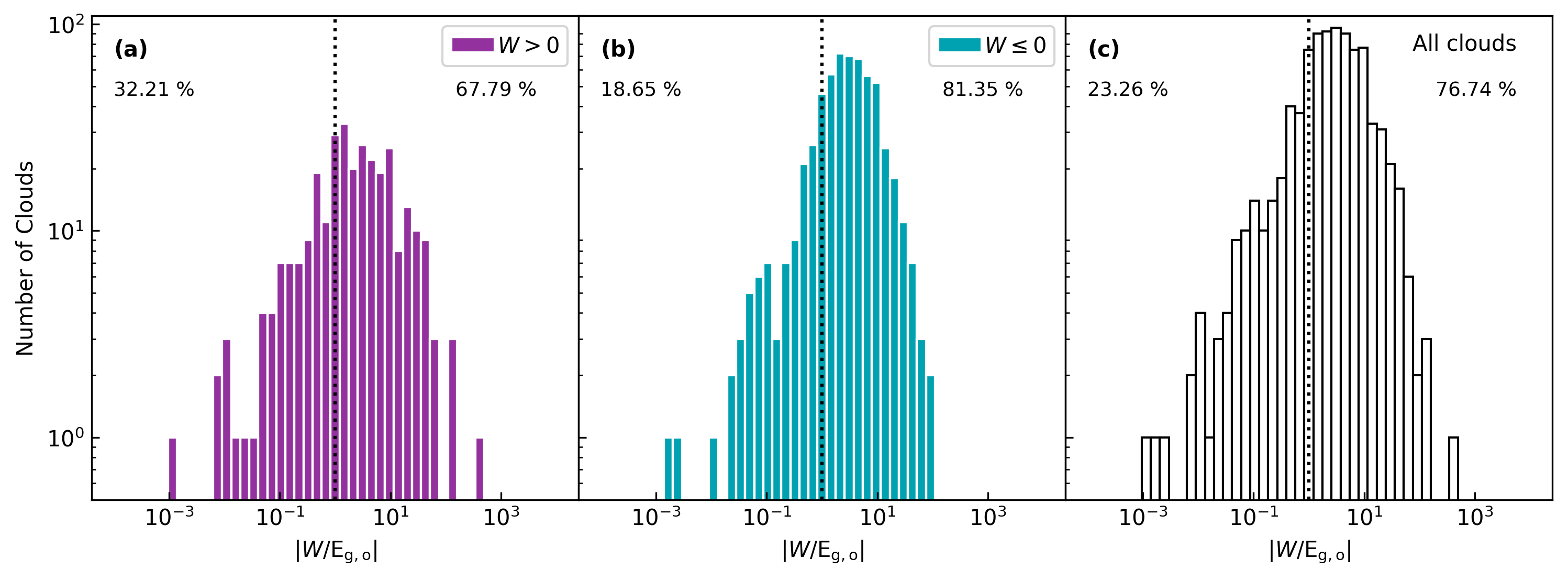} \\
  \caption{\color{black} Distributions of the ratio between the gravitational term $W$ and the energy of a spherical homogeneous cloud, $|\Egravsph| = 3GM^2/5R$. Left panel, distribution for clouds for which the gravitational term $W$ is positive, middle panel, distribution for clouds for which $W\le0$, and right panel, for all clouds. In all panels, the dotted line is located at $|W/\Egravsph|=1$. The numbers show the percentages of clouds with $|W/\Egravsph|$ larger or smaller than unity, respectively.} \label{fig:histoWoverEgsph}
}
\end{figure*}

It is clear from this figure that, statistically speaking, the magnitude of the gravitational term $W$ is systematically larger than the magnitude of the gravitational energy of a sphere with constant density $\Egravsph$, regardless of whether the gravitational term is positive or negative.   This indicates that it is necessary to estimate the gravitational content of molecular clouds in a more reliable way than just using the traditional $\Egravsph=-3GM^2/5R$, {since this quantity does not account for the external pulls and compressions that the cloud can suffer.  }

Since in Fig.~\ref{fig:histoWoverEgsph} we are plotting only the relative importance of $W$ compared to $\Egravsph$, there is no information on whether the cloud is actually bound or not. For that matter, we now turn to the virial parameter.

\subsubsection{Classical vs full virial parameter}
\label{sec:ClassicalvsFull}

In Fig.~\ref{fig:histoAlfavirClass} we show the histogram of $\alphavirclass$ for all our clouds, while in Fig.~\ref{fig:histoAlphaVirFull} we show $\alphavirfull$, separated by cases: in the left panel we include only clouds with $W>0$. In the middle panel, only clouds with $W\le0$, while in the right panel we include all clouds.  Similarly, Tables \ref{table:percentages1} and \ref{table:percentages2} and show percentages of clouds with\footnote{\color{black} Recall that in our new definition, $\alphavirfull$ can be negative, reason for which we wrote the absolute values} $|\alphavir| \le 2$ (odd columns) and $|\alphavir| > 2$ (even columns). The first table shows the values for the classical viral parameter, $\alphavirclass$, while the second table shows the values for the full virial parameter, $\alphavirfull$.  From  these figures and tables we can draw the following conclusions:

\begin{figure}
     \center{\includegraphics[width=\columnwidth]{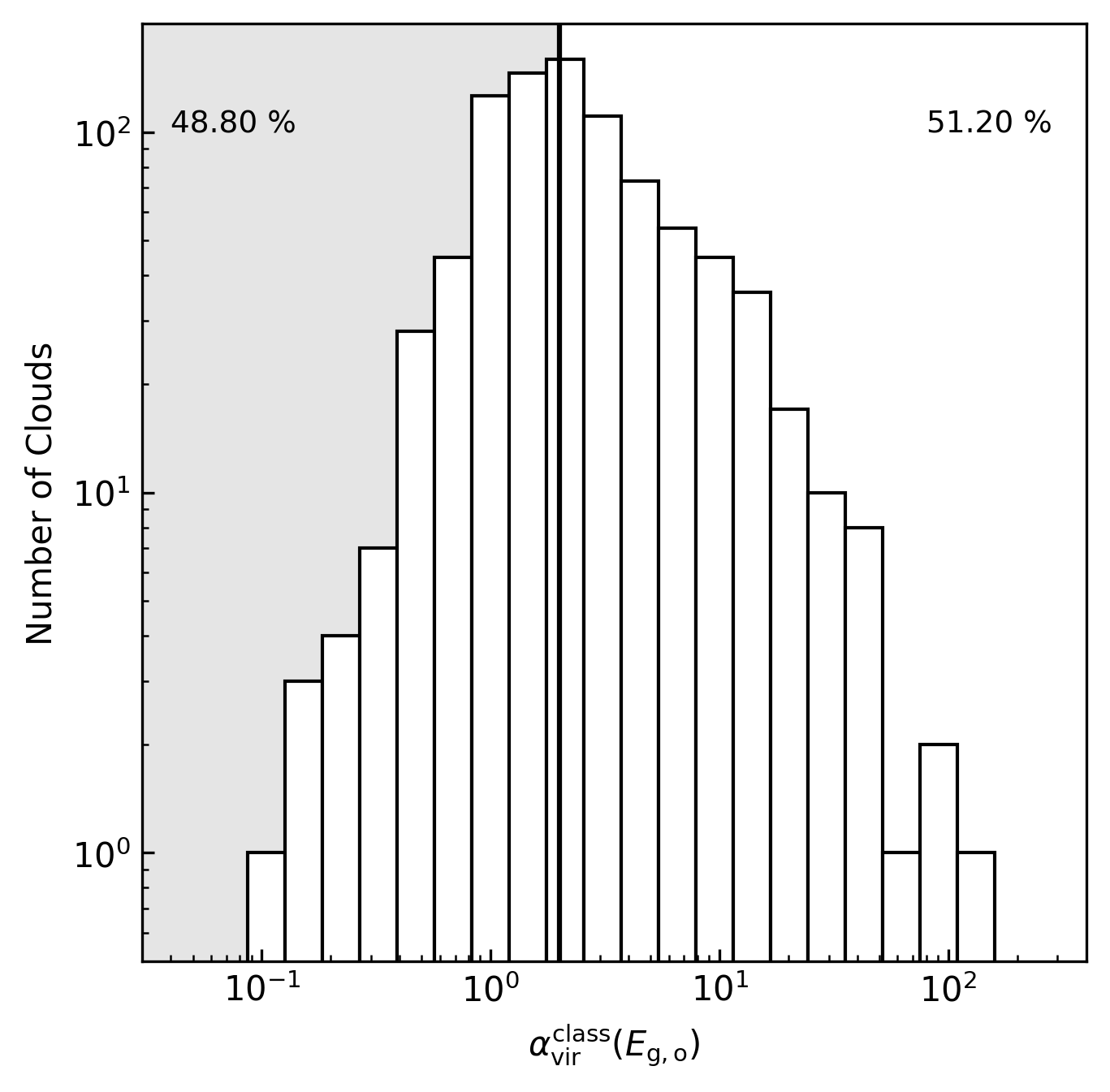} \\
  \caption{\color{black} Histogram of the classical virial parameter, $\alphavirclass$.  The vertical line at $\alphavirclass=2$ marks the division between what we believe are bound (left) and unbound (right) clouds when we use the classical virial parameter. Notice the excess of clouds at $\alphavirclass>2$ }
  \label{fig:histoAlfavirClass}
}
\end{figure}

\begin{figure*}
     \center{\includegraphics[width=2\columnwidth]{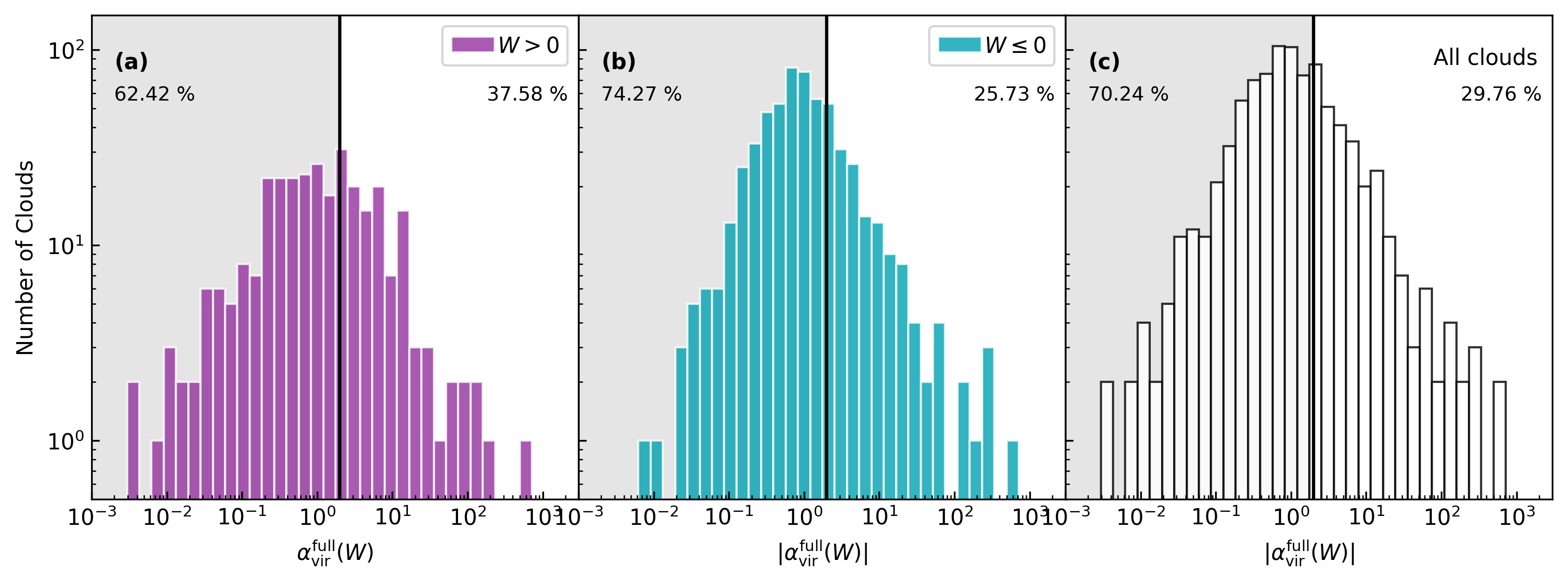} \\
  \caption{\color{black} Histograms of the full virial parameter, $\alphavirfull$. The right panel shows clouds for which $W>0$, the middle panel, $W\le0$, and the right panel all clouds. The vertical line in each panel at $\alphavirfull=2$ marks the division between bound (left) and unbound (right) clouds when we use the full virial parameter. Notice the excesses of clouds in all situations at $\alphavirfull\le2$}. \label{fig:histoAlphaVirFull}
}
\end{figure*}

\begin{enumerate}
  \item While a classical analysis will suggest that at least half of the clouds ({51.2\%}, see Fig.~\ref{fig:histoAlfavirClass} and column 2 in Table~\ref{table:percentages1}) are dominated by turbulence, the more complete view using the full virial parameter (see right panel of Fig.\ref{fig:histoAlphaVirFull}) will show that 70\%\ of clouds are actually gravity dominated, not turbulence dominated, i.e., $|\alphavirfull| < 2$ (see also columns 1 and 3 in Table~\ref{table:percentages2}). 

  \item The previous result does not mean that clouds will be necessarily gravitationally bound since the whole gravitational term $W$ can be either positive or negative. In the former case, the cloud is actually undergoing gravitational stresses that will tend to disrupt it, rather than make it collapse. In fact, as we can see from columns 1 and 2 of Table~\ref{table:percentages2} (see also left and middle panels of  Fig.~\ref{fig:histoAlphaVirFull}), {$\sim$34\%} of our population of clouds have $W>0$, implying that in 1/3 of the total population of clouds, the external gravity is playing against the self-gravity, regardless of the value of their kinetic energy.
  
  \item An interesting point worth noticing is that, from the 34\% of clouds that are been torn apart by gravity (see previous item), only in {$\sim$38\%} of them the kinetic energy actually will dominate over the disruptive gravitational energy. In other words, gravity can play a more relevant role than turbulence also in ripping apart the clouds, stressing the importance of accounting for the whole gravitational energy. 
  
  \item { A similar situation occurs with the 66\%} of clouds with negative gravity (Fig.~\ref{fig:histoAlphaVirFull}, middle panel): { a majority of clouds (74\% in this case) are dominated by gravity, and only 26\%\ are turbulence dominated.} 
  
\end{enumerate}

 In order to understand better these percentages, we now take a more detailed view of the energy budget of the clouds in our simulations.
}

\begin{table}
 \caption{Percentages of clouds gravity-dominated ($\alphavir\le2$) or turbulence-dominated ($\alphavir>2)$, according to the classical virial parameter. }
 \label{table:percentages1}
\centering
\begin{tabular}{|c|c|}
\hline
 $\alpha^{\rm vir}_{\rm class} \le 2 $   & 
 $\alpha^{\rm vir}_{\rm class} > 2 $ \\ \hline     
 48.8\%  & 51.2\%    \\
 \hline
\end{tabular}
\end{table}

\begin{table*}
 \caption{Percentages of clouds gravity-dominated  according to the full virial parameter. }
 \label{table:percentages2}
\centering
\begin{tabular}{|c|c|c|c|}
\hline
  $\alpha^{\rm vir}_{\rm full}(W>0) \le 2 $  & 
  $\alpha^{\rm vir}_{\rm full}(W>0) > 2 $  & 
  $|\alpha^{\rm vir}_{\rm full} (W\le 0)| \le 2 $ &
  $|\alpha^{\rm vir}_{\rm full}(W\le 0) | > 2 $  \\ \hline 
 21.21\%  & 12.77\%       & 49.03\%  & 16.99\%   \\ 
 \hline
\end{tabular}
\end{table*}

\subsection{Tidally tensed clouds ($W>0$)}
\label{sec:tensed}

One of our more prominent results is the existence of a population of gravitationally tensed clouds, i.e., clouds where the tidal forces overcome the internal self-gravity. This result has no counterpart in energy budget studies of molecular clouds,  {\color{black} which use the classical virial parameter $\alphavirclass$. In this particular analysis, this population is large, with {$\sim$34\% }of the total population of clouds having $W>0$, as commented in \S\ref{sec:ClassicalvsFull}.
}

{\color{black}
In order to quantify how relevant the disruptive tidal energy is, compared to the classical gravitational energy $\Egravsph$, in Fig.~\ref{fig:Wgt0_vs_Eg} we compare the gravitational term ($y$-axis) of those clouds that are torn apart (i.e., $W>0$), to the absolute value of their gravitational energy $\Egravsph$ ($x$-axis). The dotted line is the identity. We want to stress that the former quantity is intrinsically positive, while the gravitational energy is intrinsically negative. As it can be seen, although at first glance it seems that both energies span a comparable range of values, from $10^{43}$ to $10^{49}$~ergs, a closer inspection shows that $W$ spans a slightly larger dynamical range. In addition, most of the points { (81\%)} are located above the identity line. Thus, statistically speaking, the stirring gravitational term, $W>0$ exhibits systematically larger values compared to $|\Egravsph|$  (see also left panel of Fig.~\ref{fig:histoWoverEgsph}).

\begin{figure}
     \center{\includegraphics[width=\columnwidth]{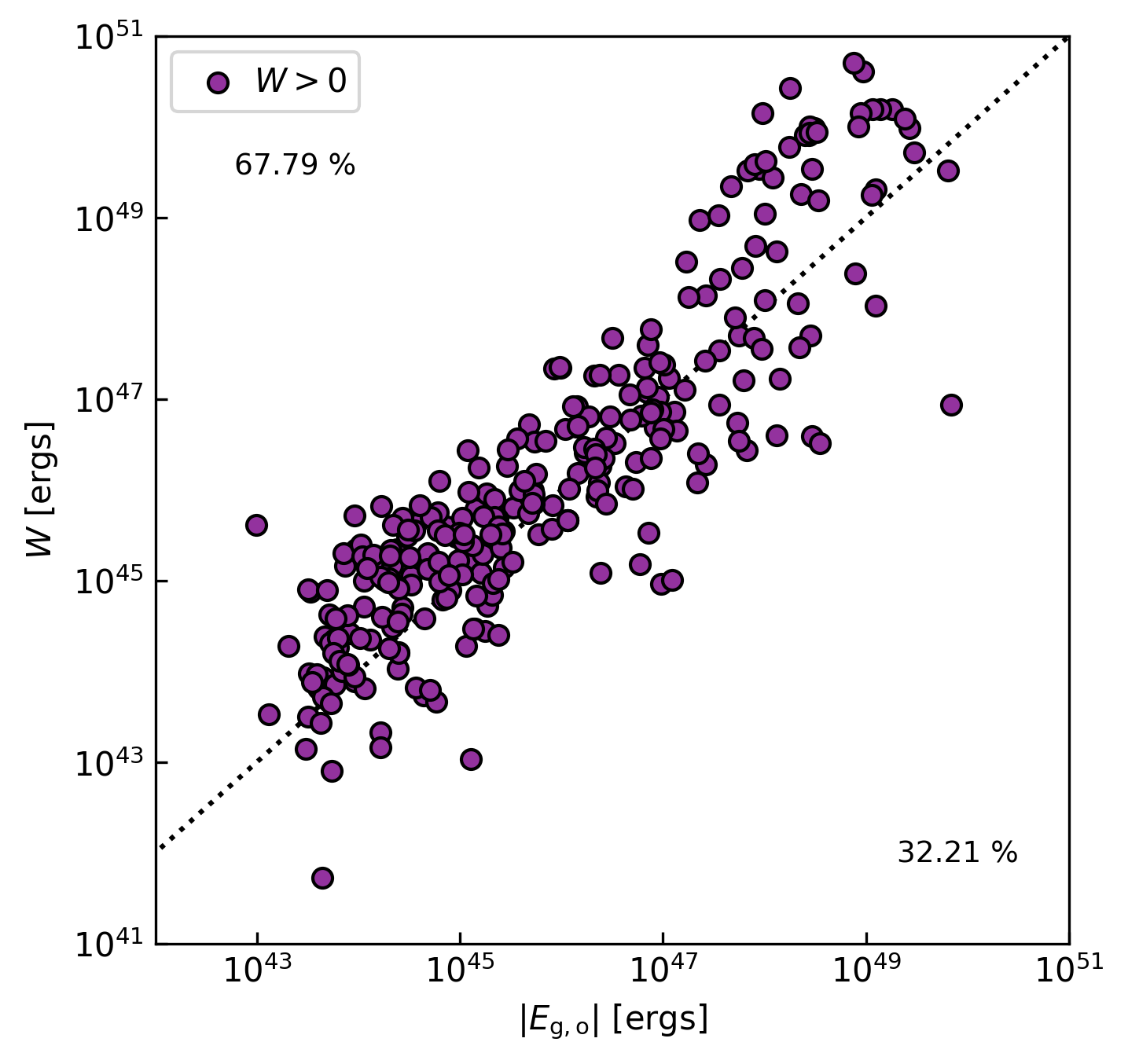} \\
  \caption{ \color{black} Gravitational term ($W$) {\it vs.} the absolute value of the gravitational energy, $\Egravsph = -3GM^2/R$ for tidally disrupted clouds ($W>0$).  The locus for which $W = |E_{\rm g}|$ is denoted by the dotted line. Note that although visually it seems there is a correspondence between $W>0$ and $\Egravsph$, these terms have substantially different meanings, since $W>0$ implies disruption, while $\Egravsph$ implies boundness. Note that, statistically speaking, $W$ is larger than $|\Egravsph|$} 
  \label{fig:Wgt0_vs_Eg}
}
\end{figure}

In Fig.~\ref{fig:alphagt0_vs_alpha} we compare the full virial parameter $\alphavirfull$ ($y$-axis) of those clouds that are tidally stirred ($W>0$), to their classical virial parameter, $\alphavirclass$ ($x$-axis). We divide this space into four regions, in order to distinguish whether both  $\alphavir$ are larger or smaller than 2. In addition, we provide percentages of clouds in the four areas, indicating, in blue, the percentage with respect to the total population of clouds, and in purple, the percentage with respect to the population of clouds shown only in this figure. 

The first point to notice from this plot is that, in order of magnitude, the full and the classical virial parameters are comparable in the sense that they span similar ranges, from 0.1 to 100, { although the span of $\alphavirfull$ is slightly larger.  Additionally, we also note that } the classical virial parameter is statistically overestimated with respect to the full virial parameter. More important, however, is the fact that according to the values of $\alphavirfull$ for gravitationally torn apart clouds ($W>0$), most of them (62\%) are { still dominated by (tidal) gravity, not by turbulence (see also Fig.~\ref{fig:histoAlphaVirFull}, left panel).} This suggests that, at least in part, molecular cloud turbulence may very well have a tidal gravitational origin, a point that we will discuss in \S\ref{sec:GravitySourceTurbulence}.

}

\begin{figure}
     \center{\includegraphics[width=\columnwidth]{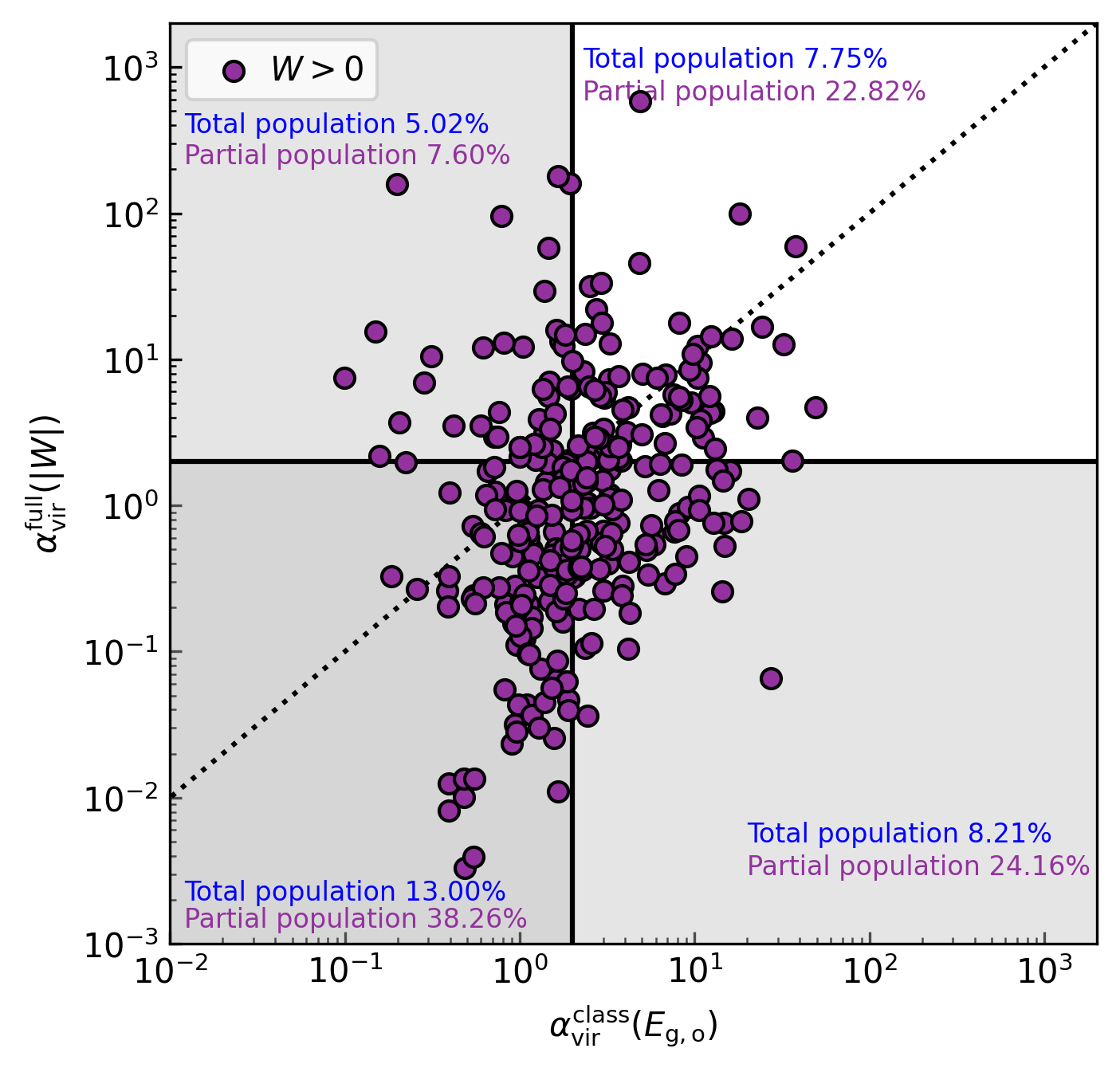}
  \caption{\color{black} Full virial parameter $\alphavirfull$ for clouds with positive gravitational energy budget, $W>0$, {\it vs.} the classical virial parameter, $\alphavirclass$. The dashed lines denote the identity line, while the solid lines denote $\alphavir=2$. 
  \label{fig:alphagt0_vs_alpha}}
  }
\end{figure}

{\color{black}

}

\subsection{Tidally compressed clouds} 
\label{sec:compressed}

\begin{figure}
     \center{\includegraphics[width=\columnwidth]{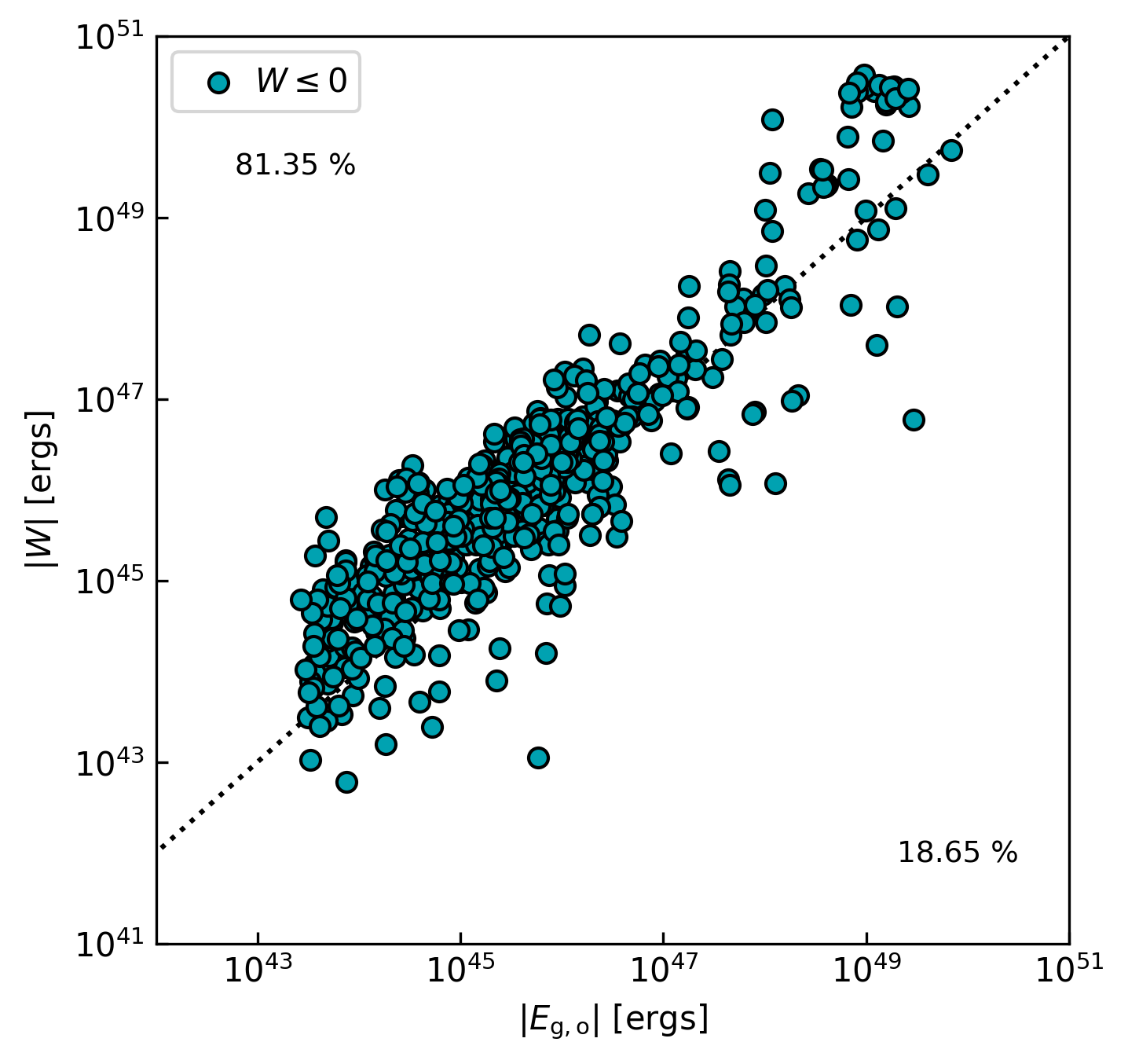}    
  \caption{ \color{black} Absolute value of the gravitational term ($W$) {\it vs.} absolute value of the gravitational energy, $\Egravsph$ (eq. [\ref{eq:Egravsph}]) for gravitationally-bound clouds ($W \le 0$). The dashed line denote the locus where $|W| = 2|E_{\rm g}|$. The locus for which $|W| = |E_{\rm g}|$ is denoted by the dotted line. Note that $|W|$ is statistically larger than $|\Egravsph|$. \label{fig:Wle0_vs_Eg}
}}
\end{figure}

We now show our results in the case where the gravitational term $W$ works as a binding energy, i.e., $W\le0$.  In Fig.~\ref{fig:Wle0_vs_Eg} we show the absolute value of the gravitational term $W$ against the absolute value of their gravitational energy, computed as if the cloud were a homogeneous sphere, $\Egravsph$. {\color{black} As in Fig.~\ref{fig:Wgt0_vs_Eg}, the dotted line represents the identity line. Again, although both energies span a similar range of values, a large majority of the population (81\%, see middle panel in Fig.~\ref{fig:histoWoverEgsph}) of clouds have $|W| > |\Egravsph|$.  This result shows that, since clouds are neither isolated nor homogeneous, the gravitational energy from an isolated homogeneous cloud is a gross approximation that tends to underestimate the actual gravitational content of clouds, also in the case in which the gravity acts as a binding energy. In other words,} gravitational tides, as well as the inner structure, enhance the gravitational binding energy of the clouds.

The tendency of $W \leq 0$ to be more negative than $\Egravsph$ suggests that the classical virial parameter implicitly assigns an over-role of turbulent motions. This is shown in Fig.~\ref{fig:alphaWle0_vs_alpha}, where we compare the absolute value of the full virial parameter $\alphavirfull$ (, eq. [\ref{eq:alphavirreal}]) of clouds with $W\le 0$ ($y-$axis), to their classical virial parameter, $\alphavirtyp$ ($x-$axis, eq. [\ref{eq:alphavirclass}]). The lines and the shaded areas have the equivalent representation as those in Fig.~\ref{fig:alphagt0_vs_alpha}. In this figure there is a clear excess of clouds below the identity line, indicating again an overestimation of $\alphavirclass$ compared to $|\alphavirfull(W\le0)|$. As a consequence of this overestimation, a typical analysis using  $\alphavirclass$ might conclude that\footnote{\color{black}{See the partial population numbers in Fig.~\ref{fig:alphagt0_vs_alpha}.}} more than one half {\color{black} ($18.13\% + 35.23\% \sim53\%$) of the clouds are on the right hand side of this figure, and thus, appear to be unbound, while the actual gravitational content indicates that only {$7.6\%+18.13\%\sim 26\%$} of the clouds are in the upper part of the plot, and  will be turbulence-dominated, (see also the middle panel of  Fig.~\ref{fig:histoAlphaVirFull} and columns 3 and 4 in Table~\ref{table:percentages2}).

\begin{figure}
     \center{\includegraphics[width=\columnwidth]{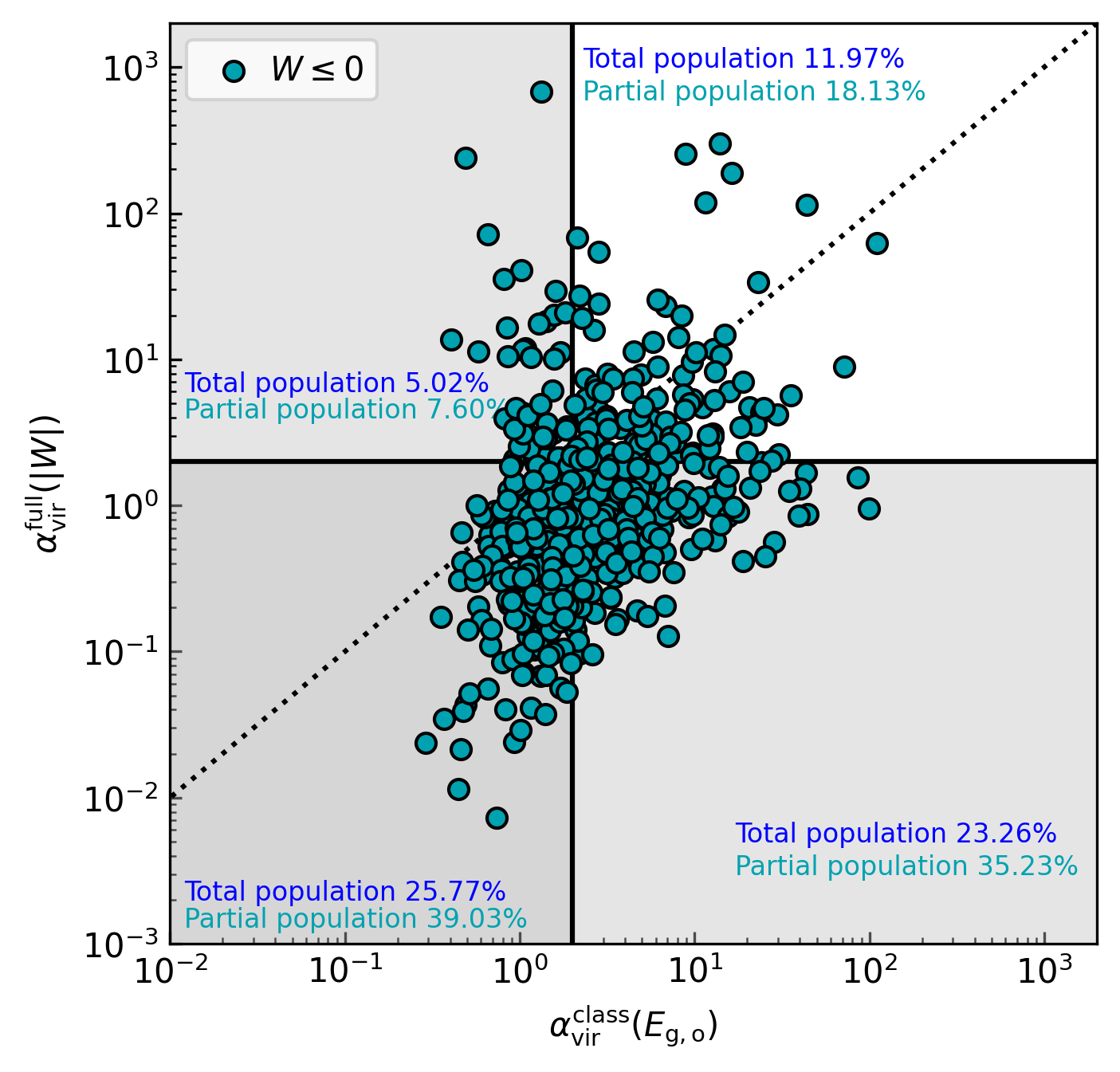} \\
  \caption{{\color{black}
  The absolute value of the full virial parameter {\it vs.} the classical virial parameter (recall that, $\alphavirfull \le 0$, since we are plotting clouds with $W\le 0$). The excess of $\alphavirclass$ over $|\alphavirfull|$ suggests that there could be a tendency to interpret bound clouds as unbound, on a classical virial parameter basis.} \label{fig:alphaWle0_vs_alpha}
}}
\end{figure}

\subsection{Virial parameter vs. mass}

In Fig.~\ref{fig:alphavir_vs_mass} we plot the virial parameters against the mass of the clouds. The panels on the left correspond to $\alphavirclass$, while the panels on the right correspond to $\alphavirfull$. The upper row corresponds to the sample containing only those clouds that have positive values of the gravitational term $W$, while the lower row, the clouds with $W\le0$. In all panels, the gray area represents the locus of clouds with $|\alphavir| < 2$. 

\begin{figure*}
 \center{
 \includegraphics[width=2\columnwidth]{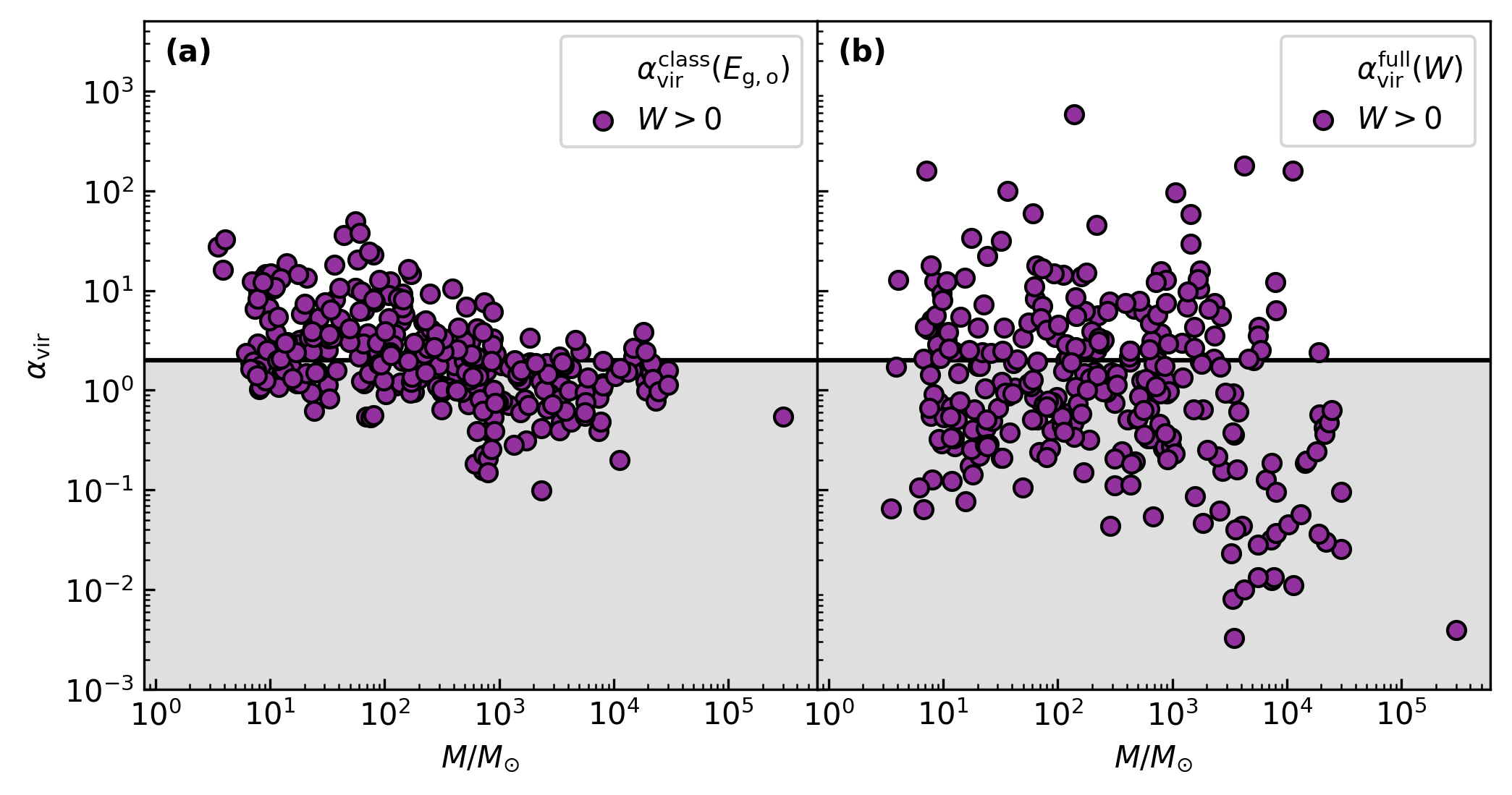}
 \includegraphics[width=2\columnwidth]{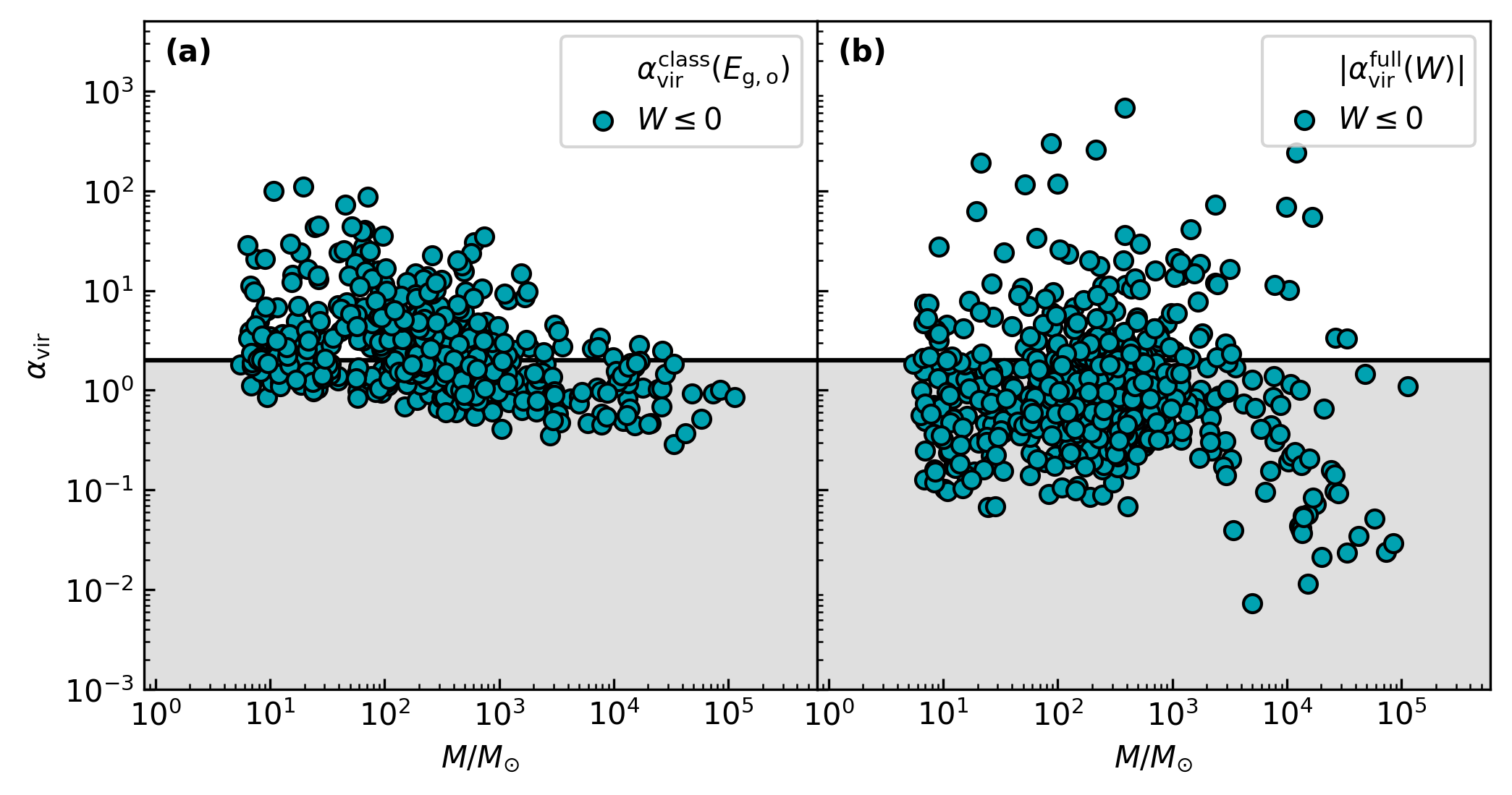}
  \caption{\color{black}Upper panels: virial parameters {\it vs.} mass for clouds with positive net gravitational energy budget, $W>0$. Lower panels: virial parameters vs mass for clouds with negative gravitational energy budget, $W\le0$. Left column: classical virial parameter, $\alphavirclass$. Right column: full virial parameter, $\alphavirfull$. Notice the lack of anticorrelation between $\alphavir$ and the mass of the clouds in all cases. \label{fig:alphavir_vs_mass}
}}
\end{figure*}

There is not much more to add to what has been said regarding the boundness of clouds in the different cases: there is an excess of clouds with $\alphavirclass>2$, but an excess of clouds with $\alphavirfull\le2$, regardless of whether $W$ is positive or negative.  Here we just stress again that with the classical virial parameter, we recover the typical results reported in previous works, with most clouds being overvirial, a minority of clouds with $\alphavirclass<1$, and, statistically speaking, more massive clouds being apparently more bound. 
}

\section{Discussion}\label{sec:discussion}

{\color{black}

We want to start our discussion by stressing that the correlations presented in the previous sections between $W$ and $\Egravsph$, or $\alphavirfull$ and $\alphavirclass$, are not remotely a one-to-one correspondence between the plotted quantities. On the one hand, we now have a term that indicates that a gravity-dominated cloud is actually torn apart, regardless of the additional kinetic energy contribution. On the other hand, even in the case of compression, the external field can provide an additional contribution to the boundness of the cloud that is missing in the classical analyses. This is a conceptual change in our idea of the role of gravity in molecular clouds, how it can affect the physical properties of the clouds, and how relevant it becomes to estimate, in a reliable way, the actual total gravitational energy in MCs in general, and in star-forming regions in particular, as we discuss in what follows. 
}

\subsection{Need for better estimations of the graviational energy of MCs}\label{sec:NeedBetterEgs}

The determination of the dynamical state of molecular clouds is a key, longstanding problem, whose solution has several implications for MC structure. For instance, in a turbulence-dominated environment, it is expected that filaments form and fragment into cores whenever they are gravitationally unstable. Conversely, in a gravity-dominated environment, as cores collapse, the surrounding material may slide into the locus of minimum gravitational potential, which are the lines that connect the cores, i.e., the filaments.

Similarly, the life-cycle of molecular clouds depends on their dynamical state. For instance, whether the star formation rate is constant or varies with time, strongly depends on the actual dynamical state of MCs. In the case of the  ONC for example, where the ages of the stars span $\sim$3~Myr, it has been considered that the region has been forming stars over $\sim$10~free-fall times (estimated with the current high-density gas), with a low and nearly constant star formation rate per free-fall time \citep[$\mathrm{SFR_{ff}}$,][]{Krumholz+19}. On the contrary, the very same $\sim$3~Myr will correspond to a roughly 1--2 free-fall timescales of the less-dense molecular cloud that might very well have undergone a global collapse \citep{VazquezSemadeni+19, Bonilla-Barroso+22}. In this case, the instantaneous SFR will have been increasing with time as collapse proceeds, as observed frequently in star-forming regions and numerical simulations \citep{Hartmann+12}. Thus, although the physical timescale  and the final efficiency of star formation (total mass in stars compared to the total mass in gas) are similar in both scenarios, the actual life-cycle of the cloud may be substantially different in each case, and thus, it becomes crucial to understand the actual physical state of molecular clouds.

One of the first models of molecular cloud dynamics suggested that, since turbulence is rapidly dissipated, molecular clouds should be in a state of global collapse \citep{GoldreichKwan74}. This idea was rapidly dismissed by \citet[][]{ZuckermanEvans74} because, in principle, such clouds should have high star formation rates. It is interesting to notice that the very same estimations by \citet[][]{ZuckermanEvans74} indicated that turbulence should be dissipated within 3~Myrs, a timescale comparable to the free-fall timescale of MCs. Thus, MC turbulence does not seem to be an effective ingredient to provide support, unless it is evenly replenished in time and space. Nevertheless, since then, the community discarded the idea of collapse and assumed that turbulence could provide support to clouds \citep[see the reviews by][and references therein]{McKee+93, Blitz93, VazquezSemadeni+00, MacLowKlessen04, elmegreenScalo04, ScaloElmegreen04, Ballesteros-Paredes+07, McKeeOstriker07, HennebelleFalgarone12, KlessenGlover16}, trading thus an efficiency problem by an equally unresolved and tough problem, the turbulence dissipation/replenishment one.  

One of the mechanisms envisaged to avoid rapid dissipation of turbulence in MCs was that turbulence could be produced by (non-dissipative) Alfvén waves. However, as shown by different authors at the end of the last century, even magnetic turbulence is rapidly dissipated \citep{MacLow+98, MacLow99, Stone+98, PadoanNordlund99}.
In addition, turbulence is a multi-scale phenomenon, and thus, while small-scale modes of turbulence could provide support to larger scales, large-scale modes will either compress and/or distort clouds within one dynamical timescale \citep{Ballesteros-Paredes+99a, Ballesteros-Paredes06}. Thus, it is not straightforward to maintain clouds against collapse for several free-fall times, unless isotropic, strong turbulence is injected at very small scales \citep{Klessen+00}. 

Numerical simulations of molecular cloud formation and their evolution, on the other hand, showed that as soon as clouds are assembled, they rapidly cool down due to the thermal instability of the diffuse gas, dropping their Jeans mass abruptly. During this process, turbulence is efficiently generated by a variety of hydrodynamical instabilities \citep[e.g., ][]{KoyamaInutsuka02, AuditHennebelle05, Heitsch+05, Heitsch+06}. However, it is not strong enough to support clouds, and they collapse as soon as they are formed \citep{VazquezSemadeni+07, Heitsch+08, HeitschHartmann08}. This lead \citet{Ballesteros-Paredes+11a} to suggest that the \citet{Larson81} scaling relation between velocity dispersion and size, $\sigma_v \propto R^{1/2}$, and its generalization, the {\it Larson's ratio} ${\cal L} \equiv \sigma_v/R^{1/2}$ {\it vs.} column density $\Sigma$ relationship, ${\cal L} \propto \Sigma$ \citep[][]{Heyer+09}, instead of being evidence that clouds are turbulent and supersonic, are evidence that they are undergoing hierarchical and chaotic gravitational collapse \citep[see also][]{IbanezMejia+16, Seifried+18}. In this scenario, the gravitational potential of irregular density structures induces non-thermal motions that appear supersonic when observed through molecular line emission. \citet[][]{Ballesteros-Paredes+11b} argued that, in contrast to the monolithic collapse scenario proposed by \citet[][]{GoldreichKwan74}, collapse occurs in a hierarchical and chaotic way, such that the amount of gas directly involved in the small, densest structures that lead directly to star formation is small compared to the total mass of the cloud. Such hierarchical collapse occurs in a variety of timescales: while the large, low-density scales collapse slowly, the small density enhancements collapse much faster, producing the stars that will afterward destroy the cloud well before the star formation efficiency becomes large. In this way, the star formation efficiency is limited by two means:  first, because the amount of gas that is at large densities is limited, and later, by the destructive effect of the stellar feedback on the cloud \citep[see][and references therein]{VazquezSemadeni+19}.

Since observations can only give us estimations of the mass, size and velocity dispersion of the cloud, estimations of the dynamical state of actual  molecular clouds invariably go through the evaluation of the virial parameter, eq.~(\ref{eq:alphavirBM}). This equation has made { two implicit assumptions. First, that turbulent motions are isotropic, and play at small scales. Second,} that the cloud can be assumed as a spheroid with constant density, such that eq.~(\ref{eq:Egravsph}) is applicable. It is argued typically that this estimation is only within a factor of $\sim$~2 from the actual gravitational content. Indeed, this is the case for triaxial ellipsoids, as shown by \citet[][]{BertoldiMcKee92}. However, for highly structured clouds, the actual gravitational energy, provided by eq.~(\ref{eq:Egravcl}), which includes the gravitational potential due to the mass outside the cloud, and the internal structure of the cloud, could be substantially different, as shown by \citet[][]{Ballesteros-Paredes+18}. These authors showed that, with the classical virial parameter, even collapsing cores can appear substantially over-virial, calling into question the typical estimations of the dynamical state of clouds. In addition, the fact that the mass outside the cloud may be playing a role via tidal interactions is also shown in numerical simulations of young stellar clusters, where an asymmetric cloud may play a role in pulling out its newborn stars \citep{Geen+18, Zamora-Avilez+19}.

{\color{black}
Judging from the actual gravitational content of molecular clouds, $W$, which substantially differs from the gravitational energy $\Egrav$, indicates that the classical virial parameter introduced by \citet{BertoldiMcKee92} cannot represent adequately the dynamics of molecular clouds. Tidal terms may substantially contribute to the total gravitational energy, either at molecular cloud core scales \citep{Ballesteros-Paredes+18}, or at molecular clouds within molecular cloud complex scales   \citep{Ballesteros-Paredes+09a, Ballesteros-Paredes+18, Liu+21}. As shown in the present contribution, while some clouds may feel a net inwards gravity larger than the gross estimation of the gravity of a spherical cloud,  other clouds may actually be torn apart by the total gravitational field. Interestingly, our results indicate that, on galactic scales, the net field providing such compressions and stresses is not the galactic field, but the very field of highly structured molecular cloud complexes and their newborn stellar clusters. Thus, it becomes necessary to start considering the environment of molecular clouds in order to have better estimations of the actual energy budget of MCs.}
 
{\color{black}

\subsection{The relevance of the galactic environment}
\label{sec:GalacticEnvironment}

Our results have another interesting puzzle: recent observational works suggest that the Galactic environment may have some influence on the dynamical state of MCs \citep[e.g., ][]{Hughes+13, Colombo+14, Faesi+18, Querejeta+19}. For instance, the latter authors have found that the fraction of dense gas in the M51 galaxy correlates with the local stellar mass surface density.  Similarly, \citet{Colombo+14} found that the properties of MCs in M51 depend on their position in the Galaxy, i.e., whether they are located in the arm or inter-arm region, whether they are up or downstream the arm, or closer and farther from the galactic center. These results suggest that the galactic gravitational potential contributes to the energy budget of MCs. However, 
we have shown that galactic tides from the bulge, stellar disk, stellar spiral arms, and dark matter halo are not relevant for the MC dynamics, a result that furthermore agrees with previous contributions \citep{Mihalas_Routhly68, Suarez+12, Jog13}. 

In other words: why spirally-aligned clouds do have properties that depend on their galaxy's environment, but theoretical estimations conclude that galactic tides are not relevant for the energy budget of molecular clouds? 

The answer to this question, we speculate, may be related to the very origin of molecular clouds: the atomic gas. As \citet{Mihalas_Routhly68} and \citet{Jog13} showed,  galactic tides from the spherical halo are relevant only at low densities, of the order of a few particles per cubic cm. These densities correspond to H~I gas clouds. MCs, on the other hand, are formed from the collisions of H~I streams \citep[see, e.g.,][]{Ballesteros-Paredes+99a, Ballesteros-Paredes+99b, Hartmann+01, MacLowKlessen04}, regardless the origin of these streams \citep{Dobbs+14}. During these compressions,  the bistable warm H~I proceeds to cool down rapidly \citep{Hennebelle_Perault99}, enhancing its density and thus, allowing for molecular cloud collapse.  Thus, the galactic tides seem to be relevant for the formation of the H~I clouds, which are the precursors of MCs.  If H~I clouds typically cannot be aligned randomly in a galactic disk because of tidal stresses \citep{Mihalas_Routhly68, Suarez+12, Jog13}, MCs will not be either aligned randomly. But apparently, HI clouds can survive if they are aligned spirally, and thus, MCs will be so too, regardless of the fact that the galactic tidal energy is not relevant to the gravitational budget of MCs. In a sense, although galactic tides are not important for molecular clouds, MCs can be thought of as the post-processed byproducts of galactic tides. 

}

\subsection{Gravity as a source of turbulence?}
\label{sec:GravitySourceTurbulence}

{\color{black} 

It has been suggested that molecular cloud turbulence could be fed by galactic dynamics, where the supersonic linewidths are the result of epicyclic motions of parcels of gas  \citet{Meidt+18}. The small fraction of energy injected into our MCs from the galactic potential suggests that this may not be the case, not at least as direct injection of turbulent motions on molecular clouds. As mentioned above, it however may be relevant for the turbulence in the more diffuse H~I clouds.  But as MCs are formed from the thermally bistable H~I streams, any non-linear behavior of the H~I streams will induce non-linear instabilities in the molecular gas that will furthermore trigger turbulent motions \citep[e.g., ][]{KoyamaInutsuka02, AuditHennebelle05, Heitsch+05, Heitsch+06}. Thus, again, in a sense, MC turbulence can be thought of as a byproduct of the galactic dynamics. 

Our results also suggest that a relevant source for MC turbulence is the total gravitational potential of the MC complexes, as well as the gravity from their young stellar clusters.  In the present work, we have seen that {$\sim$70\%} of clouds are gravity-dominated (i.e., $|\alphavirfull|\le 2$), either if they are bound (50\% of clouds with $W\le 0$) or tidally torn apart (20\% of clouds with $W>0$). MCs evolve on dynamical timescales, the gravitational potential of MCs complexes should also be changing on such timescales, injecting some energy as turbulent motions. Thus, our results suggest that, in addition to the stellar feedback, the net gravitational field of clouds and their newborn stars may also play a role in the injection of the kinetic energy of MCs.  
}

\subsection{Gravitational influence from young stellar clusters}
\label{sec:clusters}

{\color{black}
In order to estimate the energy budget of MCs, it is necessary to know their mass distribution. However, we usually do not consider how relevant is the gravity from the stellar clusters on the global budget of MCs.  Certainly, if not much mass transforms into stars, then young stellar objects should not play an important role in the gravitational budget of MCs.

{ We have seen however in Fig.~\ref{fig:AllWs} (right panel on the first row, labeled as $|W_s|$) that the sinks in our simulations could contribute in a significant way to the total gravitational budget of some molecular clouds. In principle, our sinks are only regions where the evolution of dense gas cannot be followed due to numerical limitations. They can be thought then as compact dense cores within larger MCs. However, there exists the possibility that at least some of them become actual stellar young massive clusters. If this were the case, our results will be suggesting that, insome cases, the gravitational energy of MCs could have a substantial contribution from their embedded, newly formed clusters.

As a plausible example, consider the Orion Nebula Cluster area. It is clear that in this region, the protostars are formed along the narrow dense gas filament that runs from north to south \citep[see][Fig.~14]{Megeath+12}.  In terms of their mass, the whole ONC area may have as much as $\sim$4,000~$M_\odot$ \citep{HillenbrandHartmann98} in gas, and a total content of $\sim$1,000~$M_\odot$ in stars \citep[see][and references therein]{Bonilla-Barroso+22}. Thus, the gravitational contribution of the cluster to the total gravity of the cloud may not be negligible. }

This gravitational influence can be relevant from the very moment of the formation of the cluster, to at least the first stages of the cloud dissipation due to the stellar feedback, as long as the expelled MC is not too far from their offspring stellar cluster, { and as long as the efficiency of star formation is reasonably large. Indeed, although on MC scales it is thought that the efficiency of star formation is of a few 1\%, this may not be the case for cluster-forming clouds, where the star formation efficiency can be as large as $\sim$30\%\ \citep{LadaLada03}. In this case, one can imagine that such objects, while forming stars, can have a substantially larger contribution to their gravitational budget from the stars that they are forming. }


}

\subsection{Additional thoughts}

{\color{black}

The results presented in this work are consistent with previous works in that, judging from the classical virial parameter $\alphavirclass$, many clouds {\color{black} (statistically speaking, the less massive ones)} appear to be unbound because the kinetic energy is larger than the gravitational energy.  The situation is however much more complex in reality. On the one hand, turbulent motions not necessarily will support the clouds, but they may promote collapse \citep{Ballesteros-Paredes06}. Indeed,  \citet{Vazquez-Semadeni+08, Camacho+16} and \citet{Baba+17} showed that half of the apparently unbound clouds in their simulations exhibit converging motions, and thus, they cannot prevent collapse but promote it.  

Second, it is frequently assumed that unbound clouds do have $2 < \alphavir$, and bound clouds $\alphavir < 2$. It should be noticed, however, {\color{black} that a non-equilibrium,} collapsing structure will develop virial values ($\alphavir\sim2$) in about one free-fall timescale, either in the case of a collapsing gas system \citep{VazquezSemadeni+07} or in a system of particles \citep{Noriega-Mendoza_Aguilar18}. In other words, $\alphavir\sim 2$ is the natural outcome of collapse, rather than of equilibrium.

Third, even for free-fall collapse, the terminal velocity is larger than the virial velocity \citep[see][]{Ballesteros-Paredes+11a}. In fact, as shown by \citet{Ballesteros-Paredes+18}, cold collapse of turbulent clouds produces velocity dispersions slightly larger than the virial velocity in one free-fall timescale.

The results of the present work indicate that the situation becomes even more complex since external gravity plays a role in the energy budget of MCs.  We have seen that the classical virial parameter underestimates the actual gravitational content of molecular clouds since it neglects the mass distribution external to MCs. In the present work, it becomes clear, however, that the gravitational energy from the external mass distribution is non-negligible, and thus, should be taken into account. 

{\color{black}
Finally, it should be mentioned that while the peer-review process of the present contribution was in progress, an interesting work by \citet{Ganguly+22} came up with a similar idea of evaluating the virial parameter and the tidal gravitational budget of MCs in numerical simulations of MCs at a galactic scale, but with a different approach. These authors show, however, that the gravitational tides of clouds due to the external clouds are negligible, in clear contrast with the results shown in the present contribution. We speculate which could be the reasons for this discrepancy.  

First of all, in the simulations presented by these authors, there are no spiral arm structure, and thus, large-scale, spatially correlated molecular gas is not present, making more difficult to tidally disrupt a cloud from the diffuse medium that surrounds it. 

Second, \citet{Ganguly+22} use average values of the tidal stresses in the whole cloud, while in the present work we are integrating them over the volume in the cloud. Thus, while these authors will be in practice averaging-out differences, we are adding them up. 

Third, although the calculations of \citet{Ganguly+22} show that the  gravitational acceleration due to the self-gravity in the clouds are, statistically speaking, mostly parallel to the total acceleration, that does not necessarily mean that the external acceleration is not relevant in a substantial fraction of clouds, or even, in a substantial fraction of the volume of a single cloud.  In principle, our eq.~(\ref{eq:W}) is a direct consequence of the momentum equation, and thus, it provides a detailed evaluation of the total gravitational content of MCs. 

Whether one or the other approach explains better the dynamics of MCs will be left for a further contribution. 
}

}
\section{Conclusions}\label{sec:conclusions}

{\color{black}

We have stressed the importance of evaluating the gravitational budget of molecular clouds by accounting for the total gravitational potential, considering all components of the galaxy they are embedded in, as well as their detailed inner structure and their young stellar clusters. Thus, we have made use of a state-of-the-art numerical simulation of a piece of a galactic disk to estimate the virial parameter of molecular clouds in two different ways: the classical virial parameter, which uses the gravitational energy of a sphere of constant density, as well as the full virial parameter, which uses the total gravitational potential from all the mass distribution. Our main results are simple:

\begin{enumerate}

  \item Molecular clouds in a galactic context exhibit virial values around unity, with a large scatter. 
  
  \item As in many other works, when using the classical virial parameter, there appears to be larger population of overvirial clouds.  By accounting for the whole gravitational potential, however, clouds tend to be subvirial, i.e., gravitationally dominated.

  \item Although gravitationally dominated, most of our clouds are not bound but tidally torn apart.
  
  \item To properly estimate the gravitational energy of the clouds, the total gravitational potential has to be included. The typically used (absolute value of the) gravitational energy of a constant-density sphere is just a lower limit to the gravitational content of molecular clouds. 
  
  \item Galactic tides (from the stellar bulge, stellar disk, spiral arms, and dark matter halo) are not relevant for MC energetics. However, since they appear to play a key role in defining the orientation of H~I clouds, MCs, which are formed from H~I, must be already aligned to the spiral structure. 
  
  \item The source for the tidal gravitational energy acting over  MCs is the structure of molecular cloud complexes themselves, and, probably, the presence of massive young stellar clusters. 

  \item Being relevant the tides, turbulent motions must have, at least partially, some gravitational origin, regardless of whether the clouds are collapsing or been torn apart. 

  \end{enumerate}
  
As we have mentioned, our results points towards a conceptual change in our idea of gravity, and how it can affect the physical properties of the clouds.
Although estimating the gravitational potential of the whole system may be a challenging task to accomplish observationally, it becomes clear that such task should start to be done, in order to better understand the dynamical state of molecular clouds. 
}

\section*{Acknowledgments}

The authors acknowledge the referee, Mordecai-M. Mac Low, whose thorough reading allowed us to substantially improve this manuscript.
L.R-G acknowledges the support and hospitality of IRyA's program \textit{Verano de la Investigación}, during the summer 2018, as well as support from \textit{Facultad de Ciencias Exactas y Naturales} and \textit{Relaciones Internacionales} at University of Antioquia.
L.R-G and J.B-P acknowledge UNAM-DGAPA-PAPIIT support through grant numbers IN-111-219. JBP furthermore acknowledges CONACyT through grant number {\tt 86372}. 
RJS gratefully acknowledges an STFC Ernest Rutherford fellowship (grant ST/N00485X/1). This work used the DiRAC@Durham facility managed by the Institute for Computational Cosmology on behalf of the UK STFC DiRAC HPC Facility (www.dirac.ac.uk). The equipment was funded by BEIS capital funding via STFC capital grants ST/P002293/1, ST/R002371/1, and ST/S002502/1, Durham University, and STFC operations grant ST/R000832/1. DiRAC is part of the National e-Infrastructure. V.C. acknowledges support from CONACyT grant number A1-S-54450 to Abraham Luna Castellanos (INAOE).
MZA acknowledges support from CONACYT grant number 320772.
The authors thankfully acknowledge computer resources, technical advice and support provided by the Dirección General de Cómputo y de Tecnologías de la Información y Comunicación, at UNAM, and by Laboratorio Nacional de Supercómputo del Sureste deMéxico (LNS), a member of the CONACYT network of national laboratories.
JBP also acknowledges Paris-Saclay University’s Institute Pascal for the invitation to ‘The Self-Organized Star Formation Process’ meeting, in which invaluable discussions with the participants lead to the development of the idea behind this work. 
This work has made extensive use of SAO-NASA Astrophysical Data System (ADS). 

\bibliographystyle{mnras}
\bibliography{references} 

%




\bsp	
\label{lastpage}
\end{document}